\documentclass[%
 reprint,
 amsmath,amssymb,
 aps,
]{revtex4-1}

\usepackage{graphicx}
\usepackage{dcolumn}
\usepackage{bm}
\usepackage{multirow}
\usepackage{color,soul}

\begin{document}


\title{Nodal Lines and Mapping to Mirror Chern numbers in Ca$_2$As Family}

\author{Ikuma Tateishi}
 \affiliation{Department of Physics, The University of Tokyo, Bunkyo, Tokyo 133-0033, Japan}
 \email{i.tateishi@hosi.phys.s.u-tokyo.ac.jp}
 




\date{\today}

\begin{abstract}
We study topological properties of materials called the Ca$_2$As family without spin-orbit coupling (SOC) by combining the first-principles calculation and a tight-binding model calculation. As a result of the calculation, we reveal that the Ca$_2$As family consists of one insulator phase and three nodal line phases including an intersecting nodal ring phase, though one of the phases is not found with realistic material parameters. Additionally, we discuss what kind of nontrivial topological invariants will emerge from each nodal line phase when SOC is introduced. We also find a mapping from a nodal line semimetal without SOC to a topological crystalline insulator with SOC. This mapping can be used to specify the realized topological phase from the candidates given by the previous phase classification method.
\end{abstract}

\pacs{Valid PACS appear here}
\maketitle


\section{\label{sec:intro}Introduction}
Recently, the topological insulator is energetically investigated \cite{TIRev1,TIRev2,TCI,TIRev3}. Especially easy and simple diagnostic methods to classify the topological phases have attracted much interest in this field \cite{FuKane,TIclass1,TIclass2,SBI,TQC}. One of the most famous diagnostic methods is the Fu-Kane $\mathbb{Z}_2$ index \cite{FuKane}. The Fu-Kane $\mathbb{Z}_2$ index is defined in time-reversal(TR) and inversion symmetric systems to classify TR protected topological insulator phases.
Nowadays, an advanced diagnostic method called the symmetry-based indicator \cite{SBI} is also used to classify the topological phases. The symmetry-based indicator can be used in all space group symmetries and the Fu-Kane $\mathbb{Z}_2$ index is included in it. The symmetry-based indicator is able to limit the candidates of combinations of topological invariants by counting the irreducible representations (irreps) on the high symmetry points in the momentum space and by considering a deformation to an atomic limit \cite{SBI,CF1}. 

Topological semimetal has also attracted attention in the filed of topological material science \cite{TSM1,TSM2,TSM3,nodal,TSM4,TSM5}. It is well known that a nodal line semimetal without spin-orbit coupling (SOC) is a kind of topological semimetals \cite{YKim,NLSM3,NLSM4,NLSM5,NLZ2Charge,NLHirayama}.
The nodal line semimetals without SOC are divided into two groups \cite{CF2}. One of them is a nodal line semimetal with nodes on the high-symmetry line or plane in the BZ. The other is a nodal line semimetal with nodes on the generic point in the BZ. The former is relatively easy to find by using the compatibility relation \cite{herring,BC} or by the first-principles calculation, which usually calculates the band dispersion along the high-symmetry line. On the other hand, the latter is difficult to find with a usual method, and thus a diagnostic method for them are desired. In a previous study \cite{CF2}, a diagnostic method for them has been proposed. In the method, the irreps of the occupied bands are counted to diagnose the existence of nodal line. Since the method is defined in an almost similar way as the symmetry-based indicator, a "mapping" between a nodal line semimetal and a topological insulator is also discussed in the previous study, i. e., what kind of topological insulator phase emerges from a nodal line semimetal phase when SOC is taken into account. However, for the former group of nodal line semimetals, the mapping has not understood generally, although the most of the proposed nodal line semimetals belong to the former group \cite{NLSM3,NLSM4,NLSM5,NLHirayama}. A previous study has proposed a suggestive example of the mapping for a inversion-symmetric case \cite{YKim}. It has been shown that the Fu-Kane $\mathbb{Z}_2$ index can be defined in a inversion-symmetric system without SOC and a material with non-trivial $\mathbb{Z}_2$ index must be a nodal line semimetal. It also revealed that a nodal line semimetal with non-trivial $\mathbb{Z}_2$ index is mapped to a topological insulator with the same $\mathbb{Z}_2$ index when SOC is taken into account \cite{NLSM3,NLSM4}. A recent progress extended the $\mathbb{Z}_2$ index to $\mathbb{Z}_4$ index, and revealed a new classified $\mathbb{Z}_4=2$ phase correspond to monopole-charged nodal line when SOC is neglected \cite{NLZ4}. It is also revealed that the $\mathbb{Z}_4=2$ monopole nodal line phase is mapped to the higher-order topological insulator phase \cite{HOTI,HOTIBi}. However, for more general cases with crystalline symmetries, the mapping has not been well understood.

To consider the mapping, Ca$_2$As family, $X_2Y(X=\mathrm{Ca,Sr,Ba},Y=\mathrm{Sb,As,Bi})$ has a useful feature in both of ${\bm k}$-space and real space.
In ${\bm k}$-space, Ca$_2$As family with SOC has three topological phases which consist of two different mirror protected topological crystalline insulator phases for Ca$_2$As and Sr$_2$Sb, and a trivial phase for Ca$_2$Bi \cite{CaAsFam}.
On the other hand, in the real space, it has been reported that Sr$_2$Bi, one of the Ca$_2$As family, is an electride. Here electride is a material with electrons localized in interstitial areas between nuclei \cite{Hirayama,Y2C}.
The previous study revealed that there is a strong relationship between an electride and a nodal line semimetal when SOC is neglected \cite{Hirayama,Y2C}.
Therefore, we consider that the Ca$_2$As family is a good platform to discuss the mapping between the topological invariants in topological crystalline insulators with SOC and nodal line semimetals without SOC.

In this paper, by using first-principles calculation, we study topological properties of the Ca$_2$As family without SOC. Furthermore, we derive a tight-binding model and discuss the mapping between the configuration of nodal lines and the crystalline topological invariant by introducing SOC into the tight-binding model. From the obtained mapping, we specify which phase in the candidates emerges from the nodal line phase.

It should be noted that some experimental results have shown the existence of oxygen atoms in the crystal structure of the Ca$_2$As family \cite{expO1,expO2}. 
However, in this paper, we do not consider the effect of oxygen, because our purpose is to find the mapping between two kinds of phase diagnostic methods. 

This paper is originated as follows.
In Sect.\ref{sec:material}, we introduce the crystal structure and symmetric properties of the Ca$_2$As family.
In Sect.\ref{sec:DFT}, we show the result of first-principles calculations for some materials in the Ca$_2$As family that are representative of topological phases.
In Sect.\ref{sec:tbmodel}, we derive a tight-binding model to describe the Ca$_2$As family and calculate the mirror Chern number using it. We also discuss the mapping between the existence of nodal lines in a system without SOC and the topological invariants in crystalline symmetric systems with SOC.

\section{\label{sec:material}$\mathrm{Ca}_2\mathrm{As}$ family}

\begin{figure}
    \centering
    \includegraphics[width=7cm]{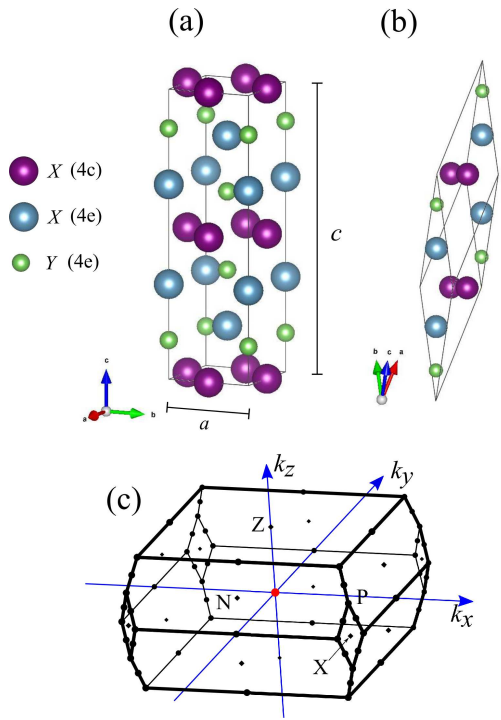}
    \caption{Crystal structure of Ca$_2$As family. Atoms in the same Wyckoff position are shown with the same color. (a) is the conventional cell picture and (b) is the primitive cell picture drawn by VESTA \cite{vesta}. (c) The Brillouin Zone of the system.}
    \label{fig:structure}
\end{figure}

Ca$_2$As family, $X_2Y(X=\mathrm{Ca,Sr,Ba},Y=\mathrm{Sb,As,Bi})$, are body-centered tetragonal crystal systems. The space group is $I4/mmm$ ($\#$139) (Fig.\ref{fig:structure}). For convenience, we introduce two different notations of the unit cell, a conventional cell (Fig.\ref{fig:structure}(a)) and a primitive cell (Fig.\ref{fig:structure}(b)). In Figs.\ref{fig:structure}(a) and (b), the atoms with the same color indicate "equivalent atoms", which means that the atoms are placed in the same Wyckoff position \cite{IntTab}. In this system, there are three nonequivalent atoms, Ca(4c), Ca(4e) and As(4e). In the conventional cell, there are twelve atoms that consist of four atoms on each nonequivalent position. On the other hand, in the primitive cell there are six atoms that consist of two atoms on each position. The primitive cell (minimal cell) is spanned by basic lattice vectors $\bm{a}_1=(a/2,-a/2,c/2)$, $\bm{a}_2=(a/2,a/2,c/2)$ and $\bm{a}_3=(-a/2,-a/2,c/2)$ (Fig.\ref{fig:structure}(b)). The lattice constants and atomic positions for all combinations are taken from crystal structure database ICSD \cite{ICSD1,ICSD2} and shown in table \ref{tab:Linfo}. The position 4e has a degree of freedom for its explicit position and it is written by the fractional coordinate in the conventional cell.
The bulk Brillouin zone (BZ) is shown in Fig.\ref{fig:structure}(c). The time-reversal invariant momenta (TRIM) are $\mathrm{\Gamma}$, Z, two X points, and four N points. Note that two P points are not TRIM but high-symmetry points.

\begin{table*}
    \centering
    \begin{tabular}{|c||c|c|c|c|c|c|c|c|c|}
    \hline
    $X_2Y$ & Ca$_2$As & Ca$_2$Sb & Ca$_2$Bi & Sr$_2$As & Sr$_2$Sb & Sr$_2$Bi & Ba$_2$As & Ba$_2$Sb & Ba$_2$Bi \\ 
    \hline \hline
    $a $ [atomic unit] & 8.75 & 9.47 & 8.92 & 9.13 & 9.45 & 9.47 & 9.69 & 9.86 & 9.95 \\ \hline
    $c/a$ & 3.36 & 3.53 & 3.50 & 3.34 & 3.48 & 3.53 & 3.38 & 3.54 & 3.55 \\ \hline
    z:$X$(4e) [fractional coord.] & 0.328 & 0.329 & 0.334 & 0.326 & 0.328 & 0.329 & 0.325 & 0.327 & 0.327 \\ \hline
    z:$Y$(4e) [fractional coord.] & 0.135 & 0.138 & 0.140 & 0.136 & 0.137 & 0.138 & 0.136 & 0.136 & 0.136 \\ \hline 
    \end{tabular}
    \caption{Parameters used in our first-principles calculation. $a$ is the lattice constant along the $a$-axis in the conventional cell and written with the atomic unit. $c$ is the lattice constant along the $c$-axis and shown as $c/a$. The $z$ coordinates of $X$(4e) atoms and $Y$(4e) atoms are shown as a fractional coordinate in the conventional cell.}
    \label{tab:Linfo}
\end{table*}

In a previous study \cite{CaAsFam}, when SOC is taken into account, Ca$_2$As and Sr$_2$Sb are suggested to be topological crystalline insulators. Particularly, Sr$_2$Sb has non-trivial $\mathbb{Z}_2$ index (weak TI) and mirror Chern number for (001) plane, while Ca$_2$As has non-trivial mirror Chern number for (1$\bar{1}$0) plane (table \ref{tab:topoinv}).

\section{\label{sec:DFT}First-principles calculation}

\begin{figure*}
    \centering
    \includegraphics[width=17cm]{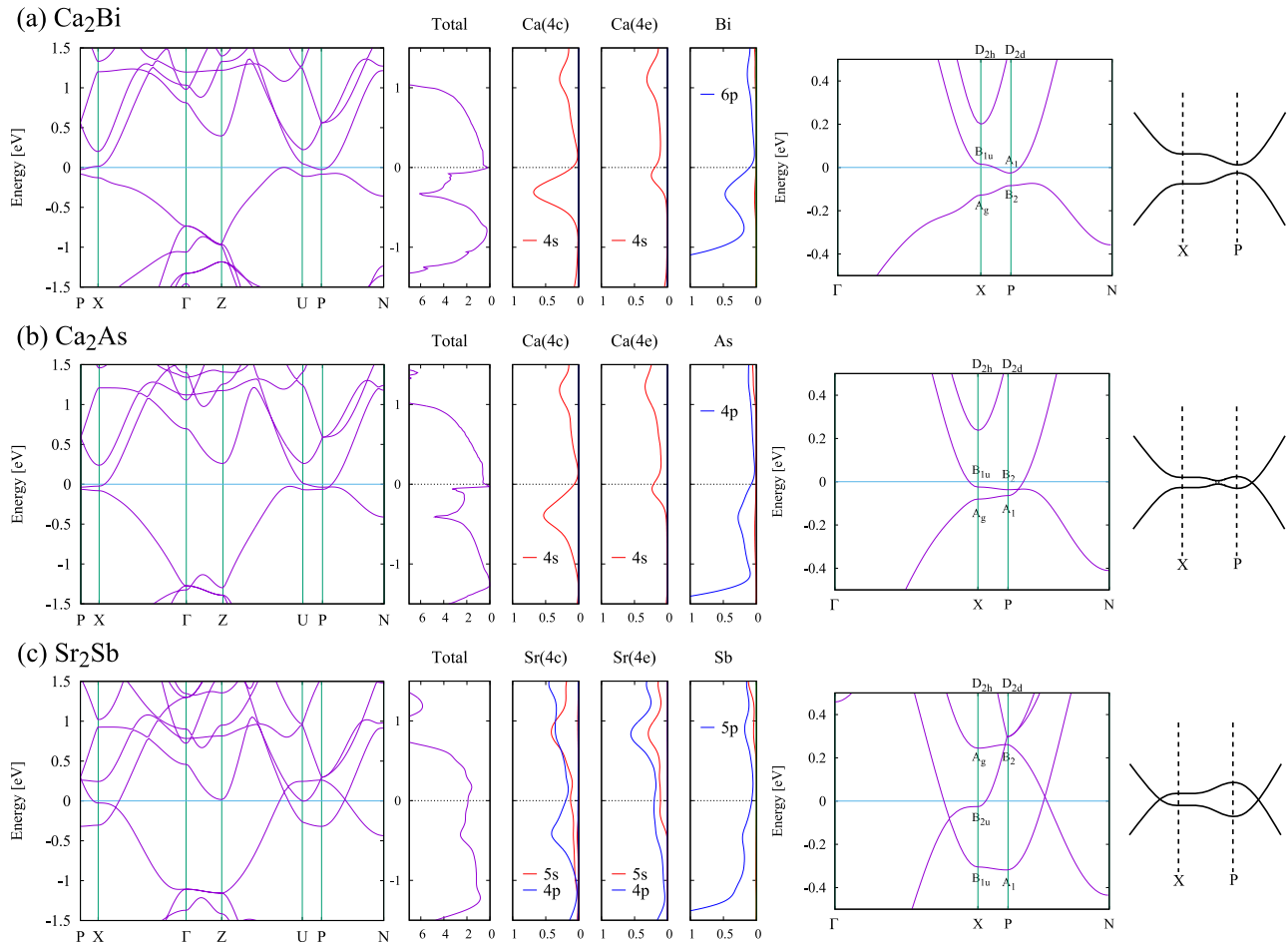}
    \caption{Band dispersions and DOS pictures without SOC for (a)Ca$_2$Bi, (b)Ca$_2$As, and (c)Sr$_2$Sb. In the panel of DOS, projected DOS to each atom and each orbital are also shown. The red lines are projected DOS to s-orbitals and the blue lines are projected DOS to p-orbitals. On the right of (a)(b)(c), a magnified picture of the band dispersion around X- and P-point is shown schematically. In the magnified band dispersion, the irreps of each band are also shown.}
    \label{fig:3bands}
\end{figure*}

In this section, we show the results of the first-principles calculation for the electronic band dispersion. We calculate all the combinations of $X_2Y(X=\mathrm{Ca,Sr,Ba},Y=\mathrm{Sb,As,Bi})$ neglecting the SOC. These calculations are performed by Quantum ESPRESSO \cite{qe}, which uses the density functional theory (DFT) \cite{dft1,dft2}. For the exchange-correlation term, generalized gradient approximation (GGA) with non-relativistic Perdew–Burke–Ernzerhof parametrization \cite{PBE} is used. The Kohn-Sham orbitals are expanded with plane waves and its cut-off energies are 40 and 170 Ry. The k-point grid on the BZ is taken as 24$\times$24$\times$24 mesh.

As shown below, we find that there are three different topological phases depending on its components (combination of atoms). Two of the three are nodal line semimetals and the other is an insulator. As representatives for those phases, three figures of the band dispersion and density of states (DOS) for Ca$_2$As, Sr$_2$Sb, and Ca$_2$Bi are shown in Fig.\ref{fig:3bands}. (The result for other materials are listed in table \ref{tab:phase} and the details are shown in appendix \ref{sec:materials}.) A magnified band dispersion and a schematic picture of the band dispersion around X- and P-points are shown on the right side of Fig.\ref{fig:3bands} for each phase. Symbols D$_{\mathrm{2d}}$ and D$_{\mathrm{2h}}$ in the magnified band dispersion represent the point groups of the little groups in the high symmetry points. For the two bands related to the band crossing, the corresponding irreps are B$_{\mathrm{1u}}$ and A$_\mathrm{g}$ for the X-point, and B$_\mathrm{2}$ and A$_\mathrm{1}$ for the P-point.
In Ca$_2$Bi (Fig.\ref{fig:3bands}(a)), no band inversion occurs around the Fermi level and the system has no node. On the other hand, in Ca$_2$As (Fig.\ref{fig:3bands}(b)), a band inversion occurs only at the P-point and a gapless node appears on the P-N line, that is a $C_2$ rotation invariant line. Since the two bands have different $C_2$ eigenvalues, this node is protected by $C_2$ rotation symmetry. Now the time-reversal (TR) symmetry is kept and the space group $\#$139 has the inversion symmetry. In the system with TR and inversion symmetry, a point node is prohibited and thus the gapless node must be a part of a line node \cite{nodal,protnode}. For this reason, we find that Ca$_2$As is a nodal line semimetal with $C_2$ rotation protected nodal lines around P-points.

Finally, in Sr$_2$Sb (Fig.\ref{fig:3bands}(c)), band inversions occur both at X- and P-points, and gapless nodes appear on the $\mathrm{\Gamma}$-X line and P-N line. In this system, a new band labeled B$_{\mathrm{2u}}$ approaches the Fermi level but it does not affect the structure of nodal lines. For the same reason explained for the case of Ca$_2$As, these gapless nodes must be parts of nodal lines. Since the $\Gamma$-X line is on a mirror invariant plane ($\sigma_z$ mirror), the node on this line is a part of a mirror protected nodal line around the X-point. The node on the P-N line, on the other hand, is a part of a $C_2$ rotation protected nodal line around the P-point. As a result, there are two different types of nodal lines in the Sr$_2$Sb system, a $C_2$ protected one and a mirror protected one.

Next, we focus on a DOS and a projected DOS shown in the middle of Fig.\ref{fig:3bands}. In Ca$_2$Bi and Ca$_2$As, the bands around the Fermi level are mainly originated from s-orbitals of Ca and p-orbitals of Bi/As. On the other hand, in Sr$_2$Sr, the p-orbitals of Sr are dominant around the Fermi level.

The charge densities of these materials also calculated and we confirmed that all materials in Ca$_2$As are electrides with typical interstitially localized electrons (See appendix D for more detail).

\section{\label{sec:tbmodel}Tight-Binding Model}
In this section, we derive a tight-binding model for the Ca$_2$As family. The difference between its components will be described by parameter tuning. By using the derived tight-binding model, we confirm the appearance of nodal lines and calculate the mirror Chern number by introducing a SOC term. From these results, we discuss a mapping between nodal line semimetals without SOC and topological crystalline insulators with SOC.

\subsection{\label{sec:tbconst}Derivation of Tight-Binding Model}

\begin{figure}
    \centering
    \includegraphics[width=6cm]{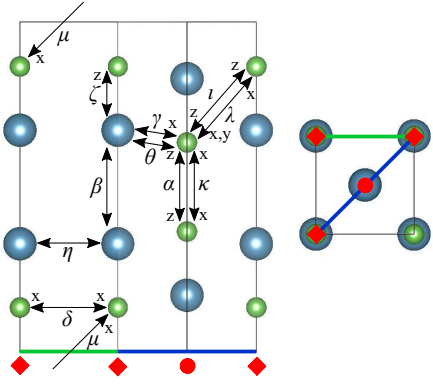}
    \caption{Hopping parameters used in the tight-binding model Eq.(\ref{eq:88tb}). The small letters, x, y and z, in the end of arrows represent the components of p-orbitals. The red square and circle correspond to the corner and center of the bottom face of conventional cell, respectively. The explicit values are shown in table \ref{tab:tbpara}.}
    \label{fig:hopping}
\end{figure}

\begin{center}
\begin{table}
    \centering
    \vspace{2mm}
    \begin{tabular}{|c||c|c|c|c|c|c|}
    \hline
     & $\alpha$ & $\beta$ & $\gamma$ & $\delta$ & $\zeta$ & $\eta^*$ \\ \hline
    [eV] & 0.485 & 0.764 & 0.331 & 0.285 & 0.270 & 0.181 \\ \hline
    \end{tabular} \\
    \vspace{2mm}
    \begin{tabular}{|c||c|c|c|c|c|}
    \hline
     & $\theta$ & $\iota$ & $\kappa$ & $\lambda$ & $\mu$ \\ \hline
    [eV] & 0.143 & 0.123 & 0.109 & 0.0918 & 0.0900 \\ \hline
    \end{tabular}
    \caption{Explicit values of hopping parameters used in the tight-binding model Eq.(\ref{eq:88tb}). The hopping channels are shown in Fig.\ref{fig:hopping}. The $*$ on $\eta^*$ means that $\eta^*$ is tuned to consider the material component dependence. $\eta=0.181$ is a parameter for the Ca$_2$Bi case.}
    \label{tab:tbpara}
\end{table}
\end{center}

First, as a low dimension model without SOC, we construct an 8$\times$8 tight-binding model using Slater-Koster's method \cite{SlaKos}, with a 4s orbital on two Ca(4e) atoms and three 4p orbitals on two As atoms. To reproduce the irreps given by the first-principles calculation, a weight of Ca(4c) must be zero. Therefore, 4s orbitals on Ca(4c) atoms are neglected here. This orbital selection seems to be incompatible with the DOS in Fig.\ref{fig:3bands}, where the projected DOS of the Ca(4c) atoms and the Sr(4c) atoms have some weight around the Fermi level. This incompatibility comes from the fact that the electrons are localized in an interstitial region. In this study, the main interest is on topological properties of the system and thus the tight-binding model should be constructed to reproduce the symmetry of each state. For this reason, using only Ca(4e) atoms is justified.

We fit the values of hopping parameters to reproduce the band dispersion by using Wannier90 \cite{wannier90}. From the result given by Wannier90, the eleven largest hopping parameters (Fig.\ref{fig:hopping}) are taken into account and others are neglected. The fitted parameters are shown in table \ref{tab:tbpara}. At the same time, the one body energies for each orbital are given as $\epsilon_s=-0.08$ [eV] for Ca(4e) 4s orbital, $\epsilon_{pz}=-2.21$ [eV] for As 4p$_z$ orbital and $\epsilon_{pxy}=-2.17$ [eV] for As 4p$_x$ 4p$_y$ orbitals. Because of the symmetry of the tetragonal system, p$_x$ and p$_y$ must have the same one body energy, whereas that of p$_z$ can be different.

The 8$\times$8 tight-binding model and its base are given as

\begin{widetext}
\begin{equation}
    H_{8\times8}(\bm{k}) = \left(
    \begin{array}{ccc}
        H_{ss:2\times2}(\bm{k}) & H_{sp1:2\times3}(\bm{k}) & H_{sp2:2\times3}(\bm{k}) \\
        H^\dagger_{sp1:2\times3}(\bm{k}) & H_{p1p1:3\times3}(\bm{k}) & H_{p1p2:3\times3}(\bm{k}) \\
        H^\dagger_{sp2:2\times3}(\bm{k}) & H^\dagger_{p1p2:3\times3}(\bm{k}) & H_{p2p2:3\times3}(\bm{k})
    \end{array}
    \right)
    \label{eq:88tb}
\end{equation}

\begin{equation}
    H_{ss:2\times2}(\bm{k}) = \left(
    \begin{array}{cc}
        \epsilon_s & -\beta-2\eta(\cos k_x + \cos k_y) \\
        -\beta-2\eta(\cos k_x + \cos k_y) & \epsilon_s \\
    \end{array}
    \right)
\end{equation}

\begin{equation}
    H_{sp1:2\times3}(\bm{k}) = \left(
    \begin{array}{ccc}
        0 & 0 & -\zeta \\
        i4\gamma e^{i\frac{c}{2}k_z}\sin\frac{k_x}{2}\cos\frac{k_y}{2} & i4\gamma e^{i\frac{c}{2}k_z}\cos\frac{k_x}{2}\sin\frac{k_y}{2} & -4\theta e^{i\frac{c}{2}k_z}\cos\frac{k_x}{2}\cos\frac{k_y}{2} \\
    \end{array}
    \right)
\end{equation}

\begin{equation}
    H_{sp2:2\times3}(\bm{k}) = \left(
    \begin{array}{ccc}
        i4\gamma e^{-i\frac{c}{2}k_z}\sin\frac{k_x}{2}\cos\frac{k_y}{2} & i4\gamma e^{-i\frac{c}{2}k_z}\cos\frac{k_x}{2}\sin\frac{k_y}{2} & -4\theta e^{-i\frac{c}{2}k_z}\cos\frac{k_x}{2}\cos\frac{k_y}{2} \\
        0 & 0 & -\zeta \\
    \end{array}
    \right)
\end{equation}

\begin{equation}
    H_{p1p1:3\times3}(\bm{k}) = H_{p2p2:3\times3}(\bm{k}) \\
    = \left(
    \begin{array}{ccc}
        \epsilon_{pxy} + 2\delta \cos k_x & 0 & 0 \\
        0 & \epsilon_{pxy} + 2\delta \cos k_y & 0 \\
        0 & 0 & \epsilon_{pz}
    \end{array}
    \right)
\end{equation}

\begin{equation}
    H_{p1p2:3\times3}(\bm{k}) = \left(
    \begin{array}{ccc}
        -\kappa e^{-ick_z} + 2\mu e^{-ick_z}\cos k_x & 0 & i4\lambda e^{-i\frac{c}{2}k_z} \sin \frac{k_x}{2} \cos \frac{ky}{2} \\
        0 & -\kappa e^{-ick_z} + 2\mu e^{-ick_z}\cos k_y & i4\lambda e^{-i\frac{c}{2}k_z} \cos \frac{k_x}{2} \sin \frac{ky}{2} \\
        i4\lambda e^{-i\frac{c}{2}k_z} \sin \frac{k_x}{2} \cos \frac{ky}{2} &
        i4\lambda e^{-i\frac{c}{2}k_z} \cos \frac{k_x}{2} \sin \frac{ky}{2} & \alpha e^{-ick_z} + 4 \iota e^{-i\frac{c}{2} k_z} \cos \frac{k_x}{2} \cos \frac{k_y}{2} \\
    \end{array}
    \right)    
\end{equation}

\begin{equation}
    \psi^\dagger = \left( 
    \begin{array}{cccccccc}
        s_{\mathrm{Ca}^1(4e)} & s_{\mathrm{Ca}^2(4e)} & p_{x:\mathrm{As}^1} & p_{y:\mathrm{As}^1} & p_{z:\mathrm{As}^1} & p_{x:\mathrm{As}^2} & p_{y:\mathrm{As}^2} & p_{z:\mathrm{As}^2} \\
    \end{array}
    \right).
    \label{eq:tb88base}
\end{equation}
\end{widetext}

\begin{figure*}
    \centering
    \includegraphics[width=16cm]{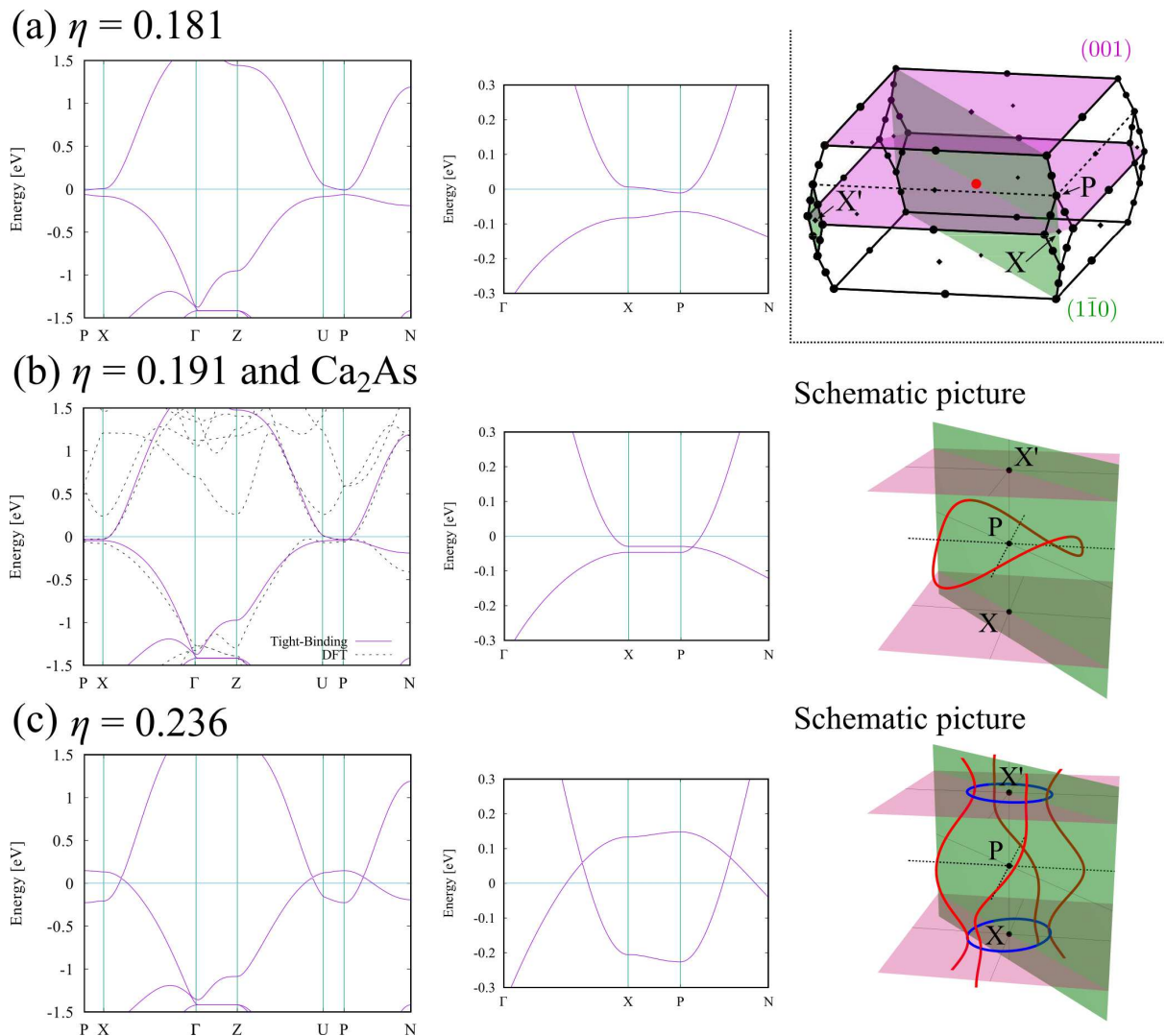}
    \caption{Band dispersions given by the tight-binding model without SOC. (a) is a phase without band inversions. (b) is a phase where a band inversion occurs only on the P-point. (c) is a phase where band inversions occur on both of the X- and P-point. In the middle panel, magnified pictures of the band dispersions are shown. On the right of (b)(c), shematic pictures of the configuration of nodal lines are shown. The green and purple planes are mirror invariant planes and the dashed lines are $C_2$ invariant lines, as shown in the BZ picture in the right top. The red lines are TR+inversion protected nodal lines and the blue ones are mirror protected nodal lines.}
    \label{fig:tbband}
\end{figure*}

The band structure of this Hamiltonian is shown in Fig.\ref{fig:tbband}. The component dependence is represented by tuning $\eta$, fixing all the other parameters as shown in table \ref{tab:tbpara} including $a=1$ and $c/a=3.36$. In Fig.\ref{fig:tbband}, the band structures for the three cases, $\eta=$0.181, 0.191 and 0.236, are shown which approximately reproduce the three systems shown in Fig.\ref{fig:3bands}. In Fig.\ref{fig:tbband}(b) for the case of $\eta=0.191$, we compare the obtained band structure with that of Ca$_2$As given by first-principles calculation (dashed line). The panels in the middle are pictures of magnified band dispersion around X- and P-point. In Fig.\ref{fig:tbband}(a), no band crossing occurs and the system has no node. In (b), a node appears on the P-N line and in (c) two nodes appear on the X-$\Gamma$ and P-N lines. These three pictures indicate that the systems of the Ca$_2$As family are well reproduced by this tight-binding model. For the case of (b) and (c), the whole configurations of the nodal lines are checked by calculating with a fine mesh in the momentum space and the results are shown in the right of Fig.\ref{fig:tbband}. The nodal lines protected with the TR and inversion symmetry are shown with red lines, and the nodal lines protected with mirror symmetry are shown with blue lines. The dashed lines (P-N lines) are $C_2$ rotation invariant lines and the green and purple planes are the mirror invariant plane characterized as $(1\bar{1}0)$ and (001), respectively. In (b), a ring of a nodal line exists around P-point and it touches the dashed lines. This nodal line oscillates in the $k_z$ direction, but keeps the $C_2$ rotation symmetry around the P-X line. In (c), in addition to the TR and inversion protected ones, mirror protected nodal lines (blue lines) appear around the X-point because the band inversion occurs at the X-point and the two bands have opposite mirror eigenvalues. On the other hand, the TR+inversion protected nodal lines (red lines) penetrate the purple planes. On the point where the red line crosses the purple plane, the mirror eigenvalues of two bands making the nodal line must be opposite because of the compatibility relation \cite{herring}. It means that this crossing point must be part of the mirror protected nodal line (blue line). In fact in the tight-binding dispersion in (c), the red line and the blue line cross with each other as shown in the schematic picture, and make intersecting nodal rings \cite{INR}. However, there is still an unproved question whether the red line always has to penetrate the purple plane if the band inversion occurs on the X-point. This point will be discussed in the next section.

\subsection{\label{sec:mapping}Phase Classification and Mapping}

\begin{table*}
    \centering
    \begin{tabular}{|c||c|c|c|c|}
        \hline
         & (A) & (B) & (C) & (D) \\ \hline \hline
        $\eta$ & $\sim 0.1884$ & $0.1884 \sim 0.1931$ & $0.1931 \sim 0.2311$ & $0.2311 \sim$ \\ \hline \hline
        \multirow{10}{*}{ \begin{tabular}{c} Nodal Lines \\ w/o SOC \\ (Schematic picture)\end{tabular} }
         & \multirow{10}{*}{ \includegraphics[width=3cm]{nodalpic0.eps} }
         & \multirow{10}{*}{ \includegraphics[width=3cm]{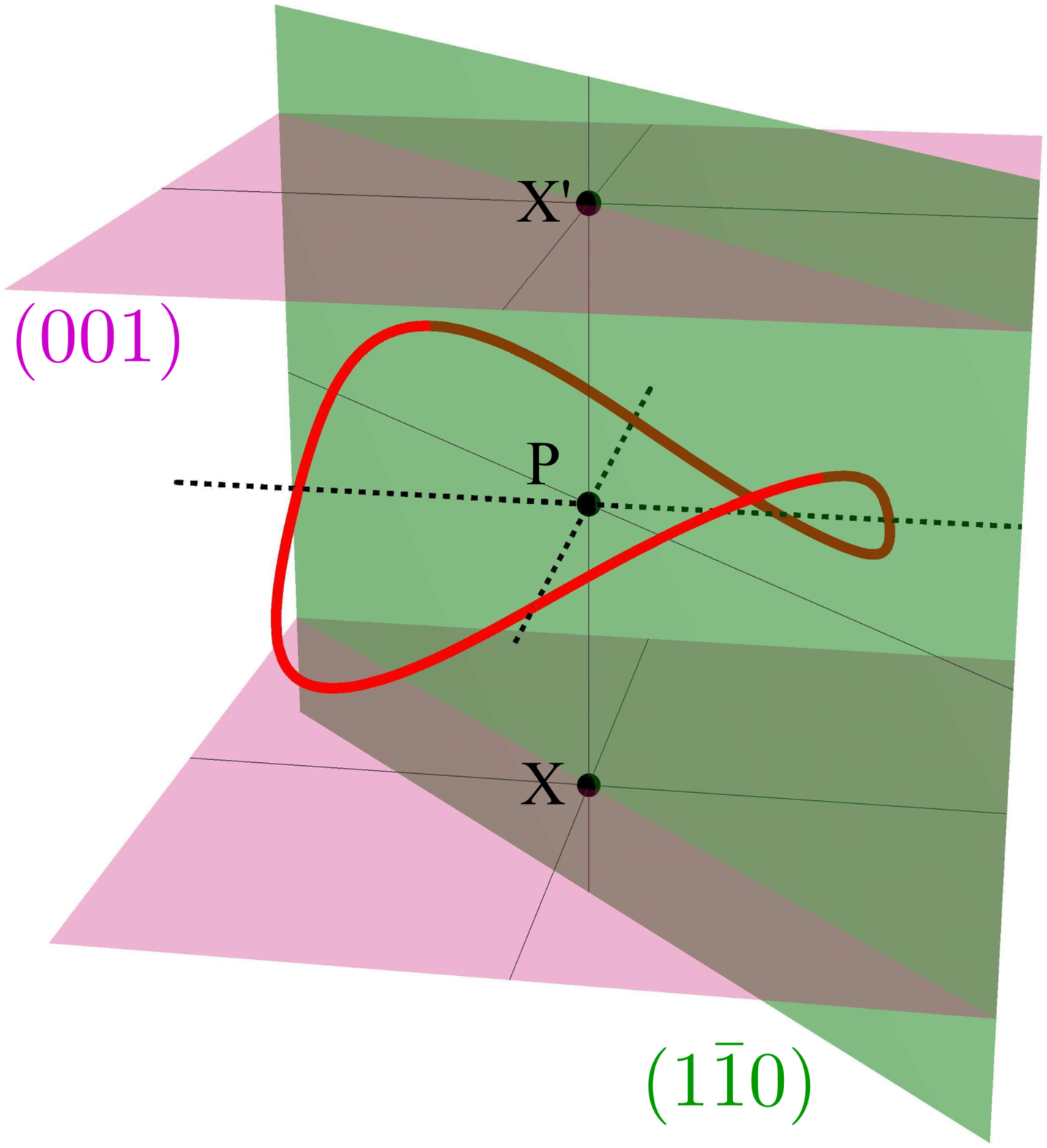} }
         & \multirow{10}{*}{ \includegraphics[width=3cm]{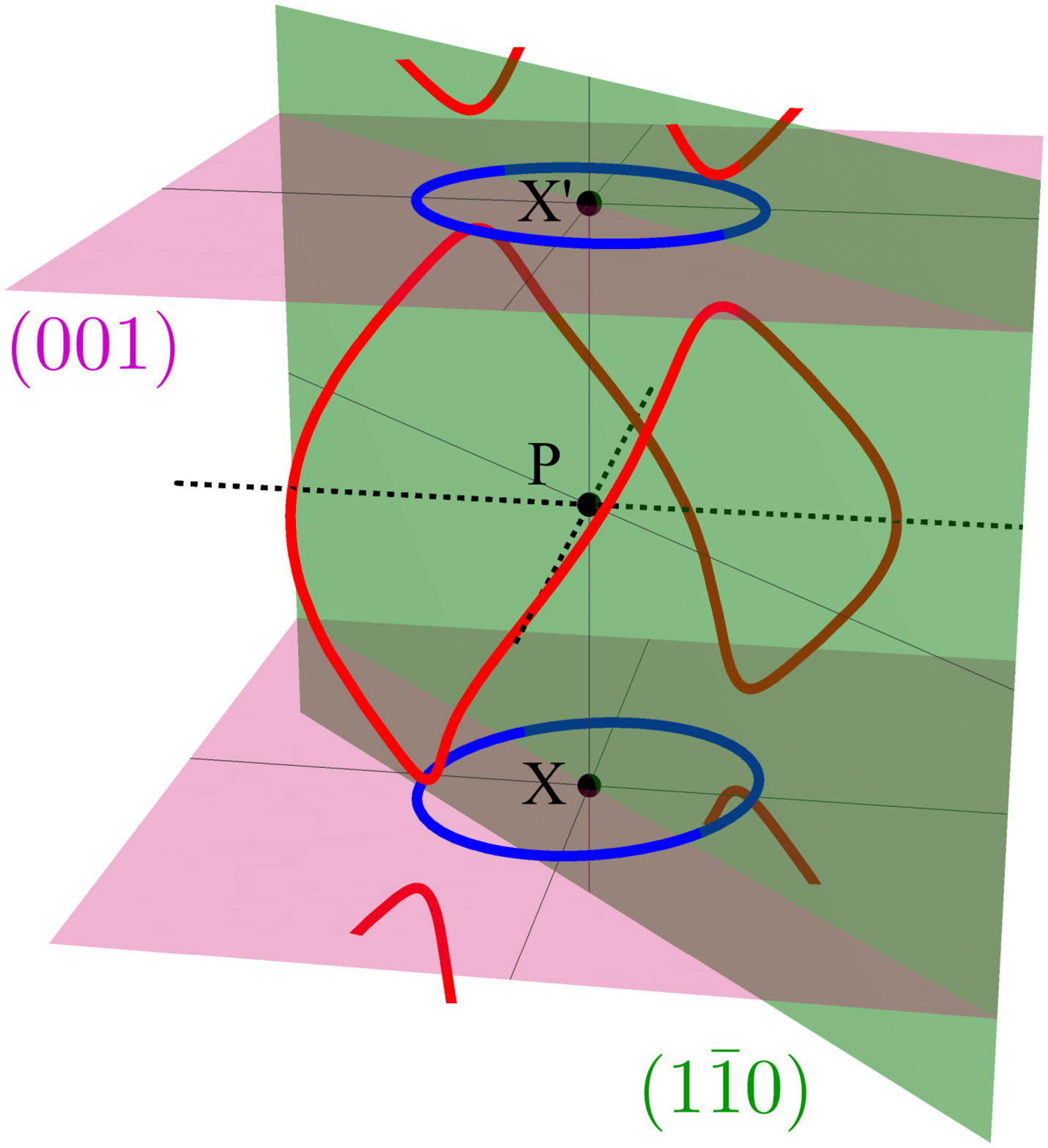} }
         & \multirow{10}{*}{ \includegraphics[width=3cm]{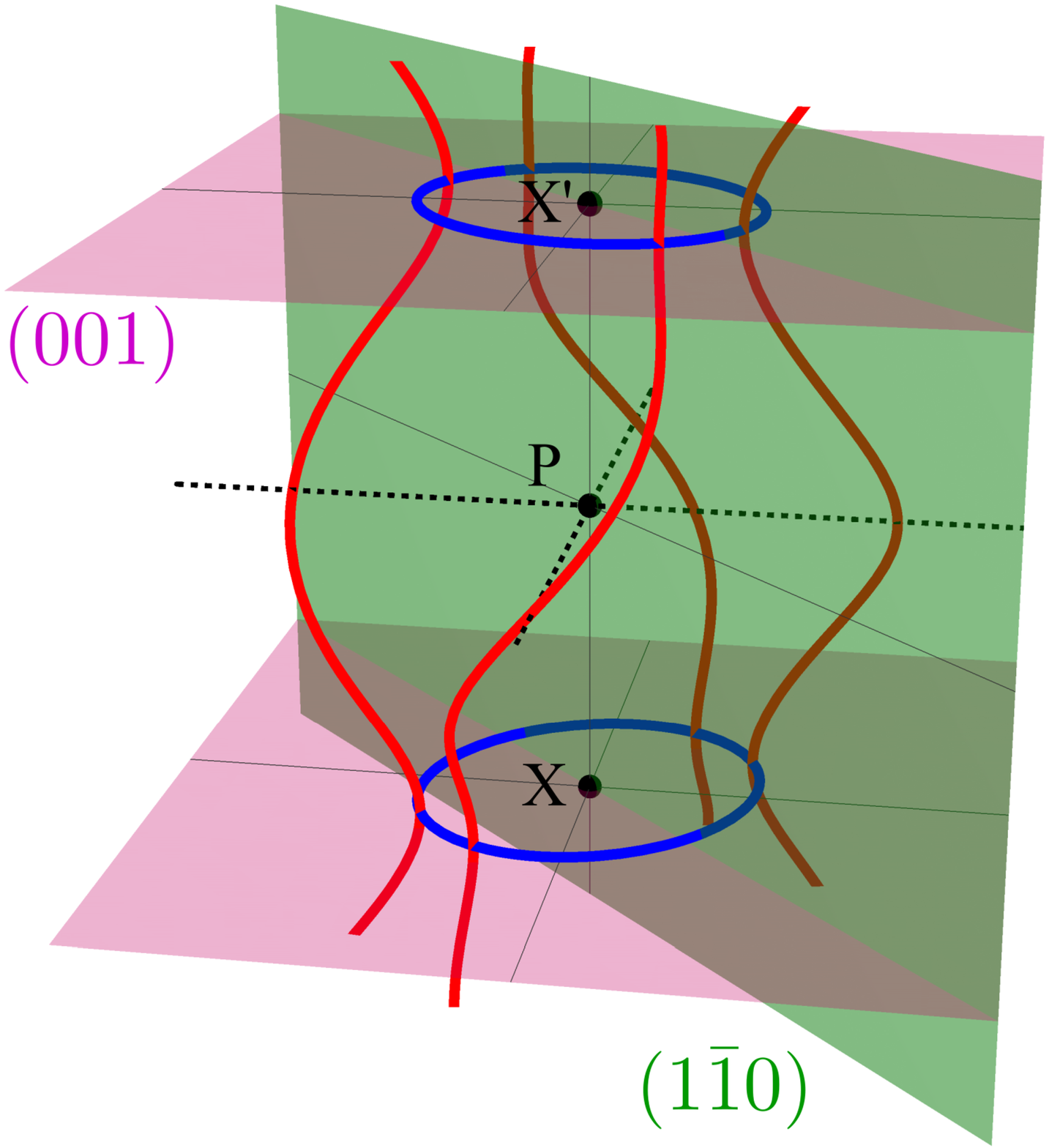} } \\
          &  &  &  &  \\
          &  &  &  &  \\
          &  &  &  &  \\
          &  &  &  &  \\
          &  &  &  &  \\
          &  &  &  &  \\
          &  &  &  &  \\
          &  &  &  &  \\
          &  &  &  &  \\ \hline
        $\mathbb{Z}_4$, $(\nu_0;\nu_1 \nu_2 \nu_3)$ & 0,(0;000) & 0,(0;000) & 2,(0;111) & 2,(0;111) \\ \hline
        \begin{tabular}{c} $({\cal Z}_2,{\cal Z}_8)$ \\ w/ SOC \end{tabular} & (0,0) & (0,4) & (1,2) & (1,2) \\ \hline
        \begin{tabular}{c} $n_{{\cal M}_{(1\bar{1}0)}}$ \\ w/ SOC \end{tabular} & 0 & 2 & 0 & 0 \\ \hline
        \begin{tabular}{c} $n_{{\cal M}_{(001)}}$ \\ w/ SOC \end{tabular} & 0 & 0 & 2 & 2 \\ \hline
        Materials & Cs$_2$Sb, Ca$_2$Bi & Ca$_2$As &  & \begin{tabular}{c} Sr$_2$As, Sr$_2$Sb, Sr$_2$Bi \\ Ba$_2$As, Ba$_2$Sb, Ba$_2$Bi \end{tabular} \\ \hline
    \end{tabular}
    \caption{Four phases given by tight-binding model with SOC. The first row shows a range of $\eta$. The second row shows a schematic picture of the configuration of nodal lines. The red lines are TR+inversion protected nodal lines and the blue lines are mirror protected nodal lines. In phase (A), there is no nodal line. The third and fourth rows show the Fu-Kane $\mathbb{Z}_2$ index and the symmetry based indicator calculated with the tight-binding model. The fifth and sixth rows show the mirror Chern numbers calculated with the tight-binding model with SOC. By checking the configuration of nodal lines, the materials in Ca$_2$As family are classified into the four phases, as given in the lowest row. }
    \label{tab:phase}
\end{table*} 

\begin{figure}
    \centering
    \includegraphics[width=8.5cm]{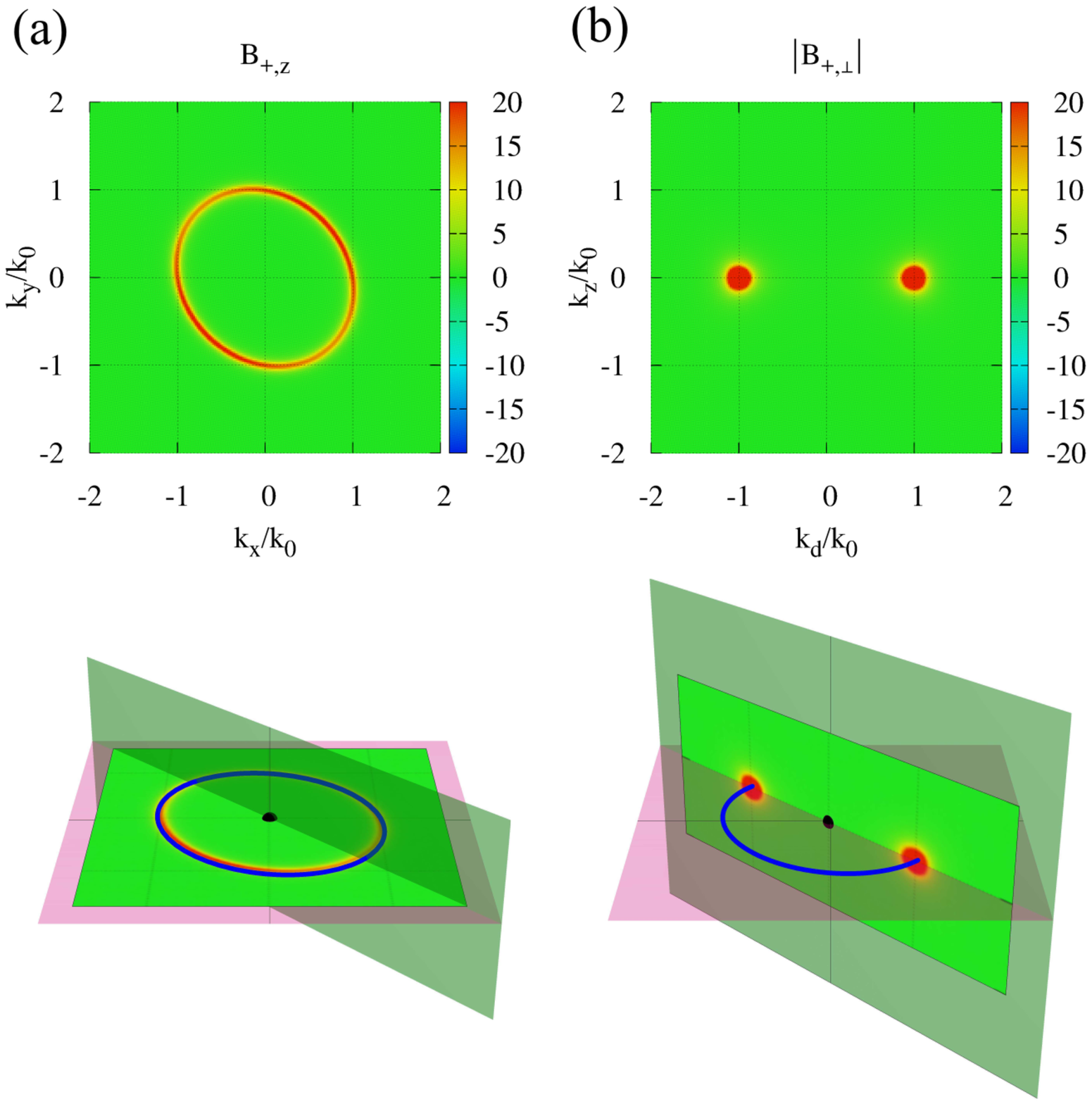}
    \caption{(a) Berry curvature on the mirror (001) invariant plane in the case where small SOC is taken into account. There is a sharp ridge on the line where the mirror protected nodal line exists when SOC is neglected. (b) The absolute value of the Berry curvature on the mirror (0$\bar{0}$1) invariant plane in the case where small SOC is taken into account. There are sharp peaks on the points where the nodal lines penetrate the mirror invariant plane when SOC is neglected. In both cases, the nodal lines can be considered as a source of the Berry curvature.}
    \label{fig:bcurv}
\end{figure}
In this section, we discuss what kind of topological phases emerge from each nodal line phase when SOC is taken into account. 

At first, the Fu-Kane $\mathbb{Z}_2$ index $(\nu_0;\nu_1 \nu_2 \nu_3)$ is easily calculated in these systems \cite{FuKane,YKim}. It is known that as long as no band inversion occurs across the Fermi level, the index does not change when the SOC is taken into account. Therefore, the Fu-Kane index can be calculated from the result without SOC.
Generally, seeing the inversion eigenvalues of two bands, one can know whether a band inversion between the two band changes the Fu-Kane index or not. In our system, the P-point is not TRIM and it has nothing to do with the Fu-Kane $\mathbb{Z}_2$ index. Therefore, the band inversion on P-point does not change the Fu-Kane index and it is $(0;000)$ for both of Fig.\ref{fig:3bands}(a) and Fig.\ref{fig:3bands}(b) case. On the other hand, the two bands on X-point, $\mathrm{A_g}$ and $\mathrm{B_{1u}}$, have different inversion eigenvalues and thus the band inversion on X-point changes the Fu-Kane index. Actually, the calculated Fu-Kane index for Fig.\ref{fig:3bands}(c) is $(0;111)$. The symmetry-based indicator $({\cal Z}_2,{\cal Z}_8)$ is also calculated with the established method \cite{SBI,CF1}. The result is shown in table \ref{tab:phase}.

Next, we consider the $\eta$ dependence of the configuration of the nodal lines. Because the nodal lines appear around P- and X-points, we can use the $\bm{k}$-$\bm{p}$ perturbation around $(k_x,k_y)=(\pi,\pi)$, keeping the $k_z$ direction periodic (See appendix \ref{app.nodal} for details of the calculation). As a result of the calculation, it is shown that the TR+inversion protected nodal line (the red nodal line in table \ref{tab:phase}(B)) appears when $\eta \simeq 0.1884$, and the mirror protected nodal line (the blue nodal line in table \ref{tab:phase}(C)) appears when $\eta \simeq 0.1931$. They touch each other when $\eta \simeq 0.2311$ table (\ref{tab:phase}(D)). By this result it is proved that there is another nodal line phase with two non-connected nodal lines, which is not found by the first-principles calculation (table \ref{tab:phase}(C)).

Next, we focus on the mirror Chern number by using the tight-binding model. By introducing a Rashba type SOC term with an amplitude $\chi$ \cite{Rashba1,Rashba2} into the tight-binding model, Berry curvatures and mirror Chern number are calculated. When SOC is taken into account in our system, the nodal line gets gaped and vanishes, and then non-trivial Berry curvature emerges. We discuss what Berry curvature emerges from the nodal line and how it contributes to the mirror Chern number (See appendices \ref{app:CNX} and \ref{app:CNP} for details of the calculation). To discuss what Berry curvature emerges from the nodal line, we calculate the Berry curvature in small SOC limit (small $\chi$ limit). We focus two mirror Chern numbers, $n_{{\cal M}_{(001)}}$ defined on the (001) mirror plane and $n_{{\cal M}_{(1\bar{1}0)}}$ defined on the (1$\bar{1}0$) mirror plane. We find that the mirror Chern numbers can be calculated as a sum of contributions from each nodal line.

First, we show in Fig.\ref{fig:bcurv} the Berry curvature which emerges from the blue nodal line around the X-point at $\bm{k}=(\pi,\pi,0)$. As shown in Fig.\ref{fig:bcurv}(a), for the (001) mirror plane, the Berry curvature has a sharp ridge on the line where the blue nodal line exists when SOC is neglected. On the other hand, for (1$\bar{1}$0) mirror plane (Fig.\ref{fig:bcurv}(b)), the Berry curvature has sharp peaks on the points where the blue nodal line penetrates the mirror plane when SOC is neglected. Integrating these Berry curvatures, we find that the blue nodal line contributes by $2$ for $n_{{\cal M}_{(001)}}$ and $-2$ for $n_{{\cal M}_{(1\bar{1}0)}}$, which are shown in the table \ref{tab:phase}(C). This result dose not change even if the red nodal line touches the blue nodal line. Therefore, this result is correct also applied to the case of table \ref{tab:phase}(D). Because the mirror Chern number is a topological invariant, it has the same value even for the large $\chi$ region. From this fact, the nodal line can be considered as a source of the mirror Chern number.

Next, we see the Berry curvature which emerges from the blue nodal line around the other X-point at $\bm{k}=(\pi,\pi,2\pi)$, which we call as X'-point. There is no difference in the calculation of $n_{{\cal M}_{(001)}}$ between the X- and X'-point. On the other hand, the sign of the contribution for $n_{{\cal M}_{(1\bar{1}0)}}$ changes from $-2$ to $2$ only on the X'-point when the red nodal line touches the blue nodal line (table \ref{tab:phase}(D)). Note that the X-point and the X'-point are connected by the $C_4$ rotation, but the (110) mirror plane is not $C_4$ invariant and thus the X- and X'-point can give different contributions to $n_{{\cal M}_{(1\bar{1}0)}}$.

Finally, we see the Berry curvature which emerges from the red nodal line around the P-point. Also in this case, the sharp peak feature like Fig.\ref{fig:bcurv}(b) is obtained, and thus the red nodal line can be considered as a source of the mirror Chern number. The contribution of the red nodal line for $n_{{\cal M}_{(1\bar{1}0)}}$ is revealed to be 2 when it penetrates the $(1\bar{1}0)$ mirror plane (table \ref{tab:phase}(B)). 

As explained before, the mirror Chern number of the system is calculated by taking a sum of the contributions form each nodal line. For $n_{{\cal M}_{(001)}}$, only the blue nodal line can contribute and thus the calculation is easy. When it does not exist (table \ref{tab:phase}(A)(B)), $n_{{\cal M}_{(001)}}=0$, and when the blue nodal line exist (table \ref{tab:phase}(C)(D)), $n_{{\cal M}_{(001)}}=2$. On the other hand, for $n_{{\cal M}_{(1\bar{1}0)}}$, we have to consider the contributions from the X-point, the X'-point, and the P-point. 

In the case of table \ref{tab:phase}(B), the red nodal line gives $2$ to $n_{{\cal M}_{(1\bar{1}0)}}$ when SOC is taken into account. In the case of table \ref{tab:phase}(C), there are also the blue nodal lines and thus the contribution from them should be considered. As shown before, The red nodal lines and blue nodal lines contribute with opposite signs to the mirror Chern number $n_{{\cal M}_{(1\bar{1}0)}}$ in the case of table \ref{tab:phase}(C). Therefore, in (C), the contributions from the red nodal line and the blue nodal line are canceled with each other, leading to $n_{{\cal M}_{(1\bar{1}0)}}=0$. 

In the case of table \ref{tab:phase}(D), only the blue nodal lines penetrate the (110) mirror plane. Considering that the nodal lines behave as sources of Berry curvature, the red nodal line dose not contribute to the mirror Chern number $n_{{\cal M}_{(1\bar{1}0)}}$. As shown before, the contribution form the blue nodal line around the X-point and the X'-point have the opposite sign in the case of table \ref{tab:phase}(D). Consequently, $n_{{\cal M}_{(1\bar{1}0)}}$ is also 0 in (D).

\begin{table}[h]
    \begin{center}
    \begin{tabular}{|c||c|c|c|c|}
        \hline
        (${\cal Z}_2$,${\cal Z}_8$) & ($\nu_0;\nu_1 \nu_2 \nu_3$) & $n_{{\cal M}_{(001)}}$ & $n_{{\cal M}_{(1\bar{1}0)}}$ & $n_{{\cal M}_{(100)}}$ \\ \hline \hline
        \multirow{4}{*}{(0,0)} & \multirow{4}{*}{(0;000)} & 0 & 0 & 0 \\ \cline{3-5}
         & & 0 & 2 & 2 \\ \cline{3-5}
         & & 4 & 0 & 2 \\ \cline{3-5}
         & & 4 & 2 & 0 \\ \hline
        \multirow{4}{*}{(0,4)} & \multirow{4}{*}{(0;000)} & 0 & 0 & 2 \\ \cline{3-5}
         & & 0 & 2 & 0 \\ \cline{3-5}
         & & 4 & 0 & 0 \\ \cline{3-5}
         & & 4 & 2 & 2 \\ \hline
        \multirow{4}{*}{(1,2)} & \multirow{4}{*}{(0;111)} & 2 & 0 & 0 \\ \cline{3-5}
         & & 2 & 2 & 2 \\ \cline{3-5}
         & & $\bar{2}$ & 0 & 2 \\ \cline{3-5}
         & & $\bar{2}$ & 2 & 0 \\ \hline
    \end{tabular}
    \caption{Examples of the candidates of topological phases for (${\cal Z}_2$,${\cal Z}_8$)=(0,0), (0,4), and (1,2). This is not all candidates but other candidates are calculated by adding each phase (See \cite{CF1} for more detail).}
    \label{tab:topoinv}
    \end{center}
\end{table}

These results show a "mapping" between where nodal lines exiting in the system without SOC and the mirror Chern numbers in the system with SOC. By using the mapping, we can indirectly know which the mirror Chern number is non-trivial by investigating the configuration of nodal lines in the system without SOC. However, as we referred to in the introduction, the symmetry-based indicator can limit the candidate of combinations of topological invariants. It is worth comparing the mapping and candidate given by the symmetry-based indicator. As explained before, the symmetry-based indicators in our system are calculated as $({\cal Z}_2,{\cal Z}_8) = (0,0)$ for (A), $({\cal Z}_2,{\cal Z}_8) = (0,4)$ for (B), and $({\cal Z}_2,{\cal Z}_8) = (1,2)$ for (C) and (D) as shown in the table \ref{tab:phase}. The explicit candidates for $({\cal Z}_2,{\cal Z}_8)=(0,0)$, $(0,4)$, and $(1,2)$ are shown in table \ref{tab:topoinv}. It is strongly limited but the candidate is not unique. Actually, by investigating the configuration of nodal lines and using the mapping, we can specify the realized phase, or decrease the candidates further. It works as follows: For the case of \ref{tab:phase}(B), $({\cal Z}_2,{\cal Z}_8)=(0,4)$. The red nodal line penetrates the $(1\bar{1}0)$ mirror plane but does not penetrate the $(001)$ mirror plane. From the mapping, the mirror Chern number should be $n_{{\cal M}_{(1\bar{1}0)}}=2$ and $n_{{\cal M}_{(001)}}=0$. The possible candidate is now unique and it is $( n_{{\cal M}_{(001)}} , n_{{\cal M}_{(1\bar{1}0)}} , n_{{\cal M}_{(100)}} ) = ( 0,2,0 )$, the second candidate of $({\cal Z}_2,{\cal Z}_8)=(0,4)$ in the table \ref{tab:topoinv}. For the case of \ref{tab:phase}(C)(D), $({\cal Z}_2,{\cal Z}_8)=(1,2)$. There are blue nodal lines on the $(001)$ mirror plane, and thus $n_{{\cal M}_{(001)}}=2$. The nodal lines exist only around X- and P-points. Therefore, no nodal line penetrates the $(100)$ mirror plane $n_{{\cal M}_{(100)}}=0$. The possible candidates are now only two and they are $( n_{{\cal M}_{(001)}} , n_{{\cal M}_{(1\bar{1}0)}} , n_{{\cal M}_{(100)}} ) = ( 2,0,0 )$ or $(2,2,0)$. As long as we just see the configuration of nodal lines, we cannot specify the realized one from the two candidates. It is because we cannot know the sign of contributions of each nodal line. For example, in the case of \ref{tab:phase}(D), $n_{{\cal M}_{(1\bar{1}0)}}=2$ when the blue nodal lines around the X- and the X'-point contribute with the same sign. On the other hand, $n_{{\cal M}_{(1\bar{1}0)}}=0$ when they contribute with the opposite sign. Although the candidate is still not unique, the number of candidates is decreased further by the mapping.

Finally, we calculated several materials of Ca$_2$As family by using the first-principles calculation and checked which phase they belong to. The band structures of these materials are shown in appendix \ref{sec:materials}. The result is shown at the bottom of the table \ref{tab:phase}. We did not find any material which belongs to (C) and all Sr$_2 Y$ and Ba$_2 Y$ belong to (D).

\section{Conclusion}
In conclusion, We calculated band structures of the Ca$_2$As family by the first-principles calculation and analyzed the topological properties on the basis of the tight-binding model. By the first-principles calculations, only three phases are found while it is revealed by a tight-binding analysis that there must be four phases. When SOC is neglected, three of the four phases are nodal line semimetals and the other is a trivial phase. We found a mapping from the nodal line phases without SOC to the topological crystalline insulator phases with SOC. We revealed that, by using the mapping, investigating the configuration of nodal lines in the system without SOC, we can specify the topological crystalline insulator phase with SOC from the several candidates given by the symmetry-based indicator. Although the result in this paper is still a case study in the Ca$_2$As family, a similar mapping is observed in other cases \cite{ITparallelFCC} and thus this mapping is presumably generalizable. More general discussions are remained as future works but this result can be an important step to a subdividing classification method.

\textit{ Note.}--- As a parallel work, the mapping is studied in the face-centered cubic system (the space group $\#$225 ($Fm\bar{3}m$)) \cite{ITparallelFCC}. In this parallel work, a similar mapping from nodal lines to mirror Chern numbers is reported. Furthermore, in FCC, a subdividing classification is given by using the mapping.
\section*{Acknowledgement}
We acknowledge the many fruitful discussions with Motoaki Hirayama, Tomonari Mizoguchi, Hiroyasu Matsuura and Masao Ogata. I.T. was supported by KAKENHI 17H02912 from JSPS and by the Japan Society for the Promotion of Science through the Program for Leading Graduate Schools (MERIT).

\bibliography{reference}
\bibliographystyle{apsrev4-2}

\clearpage
\appendix
\section{\label{app.nodal} Calculation of nodal lines}
To make the calculation easier, I derived a two by two model for the system without SOC.  The bases of the model $\psi_+$ and $\psi_-$ are the eigenstates of Eq.(\ref{eq:88tb}) for the two bands around the Fermi level on X-point, which are given by numerical calculation as

\begin{equation}
    \begin{split}
        \psi_+ = (-A,A,0,0,B,0,0,B), \\
        \psi_- = (C,C,0,0,-D,0,0,D),
    \end{split}
\end{equation}
\begin{equation*}
    A \simeq 0.6975 ~,~ B \simeq 0.1163 ~,~ C \simeq 0.7030 ~,~ D \simeq 0.07590
\end{equation*}

with the definition in Eq.(\ref{eq:tb88base}). The $\psi_+$ is an anti-bonding like state and the $\psi_-$ is like a bonding state between two Ca 4s orbitals. The two by two model is given as

\begin{equation}
    \begin{split}
        H_{2 \times 2}(\bm{k}) &= \left(
        \begin{array}{cc}
            \langle \psi_+ | H_{8 \times 8}(\bm{k}) | \psi_+ \rangle & \langle \psi_+ | H_{8 \times 8}(\bm{k}) | \psi_- \rangle \\
            \langle \psi_- | H_{8 \times 8}(\bm{k}) | \psi_+ \rangle & \langle \psi_- | H_{8 \times 8}(\bm{k}) | \psi_- \rangle 
        \end{array}
        \right), \\
        &= Z \sigma_z + Y \sigma_y  ~~(+ W \sigma_0),
    \end{split}
    \label{eq:tb22}
\end{equation}
\begin{equation}
    Z = J + K (k_x^2 + k_y^2) + L \cos c k_z + M k_x k_y \cos \frac{c k_z}{2} ,
\end{equation}
\begin{equation}
    Y = S \sin c k_z + T k_x k_y \sin \frac{c k_z}{2} ,
\end{equation}
\begin{equation}
    \begin{split}
        J =& (A^2-C^2)\epsilon_s + (A^2+C^2)\beta + 2(AB-CD)\zeta \\
        &+ (B^2-D^2)\epsilon_{pz} - 4(A^2+C^2)\eta \\
        \simeq& 0.7483 - 3.9229 \times \eta ,
    \end{split}
\end{equation}
\begin{equation}
    \begin{split}
        K =& (A^2+C^2) \eta \\
        \simeq& 0.1962 ,
    \end{split}
\end{equation}
\begin{equation}
    \begin{split}
        L =& (B^2+D^2) \alpha \\
        \simeq& 0.009357 ,
    \end{split}
\end{equation}
\begin{equation}
    \begin{split}
        M =& 2(AB+CD) \theta + (B^2 + D^2) \iota \\
        \simeq& 0.04060 ,
    \end{split}
\end{equation}
\begin{equation}
    \begin{split}
        S =& 2BD \alpha \\
        \simeq& 0.008565 ,
    \end{split}
\end{equation}
\begin{equation}
    \begin{split}
        T =& 2(AD+BC)\theta + 2BD \iota \\
        \simeq& 0.04047 ,
    \end{split},
\end{equation}
where $\sigma_y$ and $\sigma_z$ are the Pauli matrices and $\sigma_0$ is a two-dimension identity matrix, $(k_x,k_y)$ is redefined as a relative coordinate from $(\pi,\pi)$. The detail of the $W$ is omitted because the coefficient of $\sigma_0$ does not matter on the structure of nodal lines and topological properties of this model.

First, we discuss the presence of nodal lines and its $\eta$ dependence. Energy eigenvalues of the model Eq.\ref{eq:tb22} are written as $E_{\pm}= \pm \sqrt{Z^2+Y^2} ~~(+W)$. Nodal lines appear on the point where $\bm{k}$ satisfies $Z=0$ and $Y=0$. Considering $0 < M < K/2$, the condition for the existence of a solution to $Z=0$ is $J + L \cos c k_z < 0$. The sign of $J + L \cos c k_z$ on the planes of $k_z = 0$ (including the X-point) and $c k_z = \pi$ (including the P-point) are shown below.
\begin{tabular}{|c||c|c|c|c|c|}
\hline
    $\eta$ & (small) & $\simeq 0.1884$ & & $\simeq 0.1931$ & (large) \\
\hline \hline
    $J+L$ ($k_z=0$) & + & + & + & 0 & - \\
\hline
    $J-L$ ($k_z=\pi$) & + & 0 & - & - & - \\
\hline
\end{tabular}
Regarding $Y$, for $c k_z = \{0,2\pi \}$, $Y=0$ is satisfied on all $(k_x,k_y)$ and for other $k_z$ the solution to $Y=0$ is written as $k_x k_y = \frac{2S}{T} \cos \frac{c k_z}{2}$. 
The overlap of the solutions to $Z=0$ and $Y=0$ is the point where nodal lines appear. For the $c k_z = \{0, 2\pi \}$ plane, the solution to $Z=0$ makes a nodal line and it exist when $J+L<0$ is satisfied. This nodal line appears on mirror invariant plane as an ellipse and corresponds to the blue ring in Fig.\ref{fig:tbband}. For the $c k_z = \{-\pi, \pi \}$ plane, the equation $Y=0$ is written as $k_x k_y =0 $. Considering the solution to $Z=0$ is a circle on $(k_x,k_y,\pi/c)$ plane if $J-L<0$ is satisfied, nodes appear at two points on each of the $k_x=0$ line and the $k_y=0$ line. These nodes are parts of nodal lines and correspond to the nodes protected by rotation symmetry on the P-N line in Fig.\ref{fig:tbband}. For the $c k_z = \pi + \Delta$ plane, which is the plane shifted by infinitely small $\Delta$ from $c k_z = \pi$ plane, the nodes appear on the points where $(k_x,k_y)$ satisfy
\begin{equation}
    \begin{split}
        k_x k_y &= \frac{S}{T} \Delta ,\\
        0 &= J - L + {\cal O}(\Delta^2) + K (k_x^2+k_y^2) .
    \end{split}
\end{equation}
From this, it is shown that the nodes on the $c k_z = \pi$ plane are always connected to nodes on $c k_z = \pi + \Delta$ plane and they make nodal lines. For the $c k_z = \Delta$ plane, which is the plane shifted by infinitely small $\Delta$ from $c k_z = 0$ plane, the nodes appear on the points where $(k_x,k_y)$ satisfy
\begin{equation}
    \begin{split}
        k_x k_y &= -\frac{2S}{T} + {\cal O}(\Delta^2) ,\\
        0 &= J - L - \frac{2SM}{T} + {\cal O}(\Delta^2) + K (k_x^2+k_y^2) .
    \end{split}
\end{equation}
The solution to this equation exists when
\begin{equation}
    0> J + L - \frac{2SM}{T} + \frac{4SK}{T}
\end{equation}
is satisfied. It can be written as $\eta>0.2311$. Combining with the condition for the $k_z = 0$ plane, when $0.1931<\eta<0.2311$, nodal lines appear around both of the X- and the P-points but they are not connected. The configuration of nodal lines for each $\eta$ region is summarized in table \ref{tab:phase}.

\section{\label{app:CNX} Calculation of the mirror Chern number (X-point)}

Next, we introduce SOC in the tight-binding model and discuss what topological phases emerge from each nodal line phase. Considering the symmetry of bases, SOC is introduced as a Rashba type term $\chi \sigma_x (-k_y s_x + k_x s_y)$ \cite{Rashba1,Rashba2}, where the $\chi$ is an amplitude of SOC and the $s_x$ and $s_y$ are the Pauli matrices for the spin degree of freedom. The tight-binding model is now a four by four matrix written as
\begin{equation}
    H_{4 \times 4}(\bm{k}) = H_{2 \times 2}(\bm{k}) s_0 + \chi \sigma_x (-k_y s_x +k_x s_y) ,
\end{equation}
where $s_0$ is a two by two identity matrix is the spin component.

Let us focus on the mirror Chern number ${\cal M}_{(001)}$ on the $k_z = 0$ plane. On the $k_z=0$ plane, the tight-binding model is written as
\begin{equation}
    H_{4 \times 4}(\bm{k})= \left(
    \begin{array}{cccc}
        Z & 0 & 0 & \bar{U} \\
        0 & -Z & \bar{U} & 0 \\
        0 & U & Z & 0 \\
        U & 0 & 0 & -Z
    \end{array}
    \right),
    \label{eq:socX001}
\end{equation}
\begin{equation}
    \begin{split}
        U = \chi (- k_y + i k_x) , \\
        \bar{U} = \chi (- k_y - i k_x).
    \end{split}
\end{equation}
The representation of the mirror operation $(x,y,z) \to (x,y,-z)$ is now
\begin{equation}
    -i \sigma_z s_z = \left(
    \begin{array}{cccc}
        -i & 0 & 0 & 0 \\
        0 & i & 0 & 0 \\
        0 & 0 & i & 0 \\
        0 & 0 & 0 & -i 
    \end{array}
    \right).
\end{equation}
It is seen that the matrix Eq.(\ref{eq:socX001}) is block diagonalized and each block corresponds to the components with the mirror eigenvalues $+i$ and $-i$. They are separately written as
\begin{equation}
    H_+ (\bm{k}) = X_+ \sigma_x + Y_+ \sigma_y + Z_+ \sigma_z ,
\end{equation}
\begin{equation}
    \begin{split}
        X_+ &= -\chi k_y , \\
        Y_+ &= \chi k_x , \\
        Z_+ &= -Z_{k_z =0} ,
    \end{split}
\end{equation}
\begin{equation}
    H_- (\bm{k}) = X_- \sigma_x + Y_- \sigma_y + Z_- \sigma_z ,
\end{equation}
\begin{equation}
    \begin{split}
        X_- &= -\chi k_y , \\
        Y_- &= \chi k_x , \\
        Z_- &= Z_{(k_z=0)} .
    \end{split}
\end{equation}

The Berry connection for the occupied band of $H_+(\bm{k})$ is
\begin{equation}
    \begin{split}
        A_{+,x} = \frac{\chi X_+}{2R_+(R_+ + Z_+)} = \frac{-\chi^2 k_y}{2R_+(R_+ + Z_+)} ,\\
        A_{+,y} = \frac{\chi Y_+}{2R_+(R_+ + Z_+)} = \frac{\chi^2 k_x}{2R_+(R_+ + Z_+)},
    \end{split}
    \label{eq:bconnX001}
\end{equation}
\begin{equation}
    R_+ = \sqrt{X_+^2 + Y_+^2 + Z_+^2}.
\end{equation}
By integrating this along a circle with a radius $k/\chi$ in the $(k_x,k_y)$ plane, I get
\begin{equation}
    \begin{split}
            \oint d \bm{k} \cdot \bm{A} = \int_0^{2\pi} d \theta \frac{k^2}{2R_+(R_+ + Z_+)}.
    \end{split}
\end{equation}
In $k/ \chi \to \infty$ limit ,
\begin{equation}
    \begin{split}
        R_+ &\to |Z_+| , \\
        R_+ + Z_+ &\to \frac{1}{2} \frac{1}{|Z_+|} ,
    \end{split}
\end{equation}
thus the Chern number of the $+i$ component $N_+$ is calculated as
\begin{equation}
    \begin{split}
        N_+ = \oint d \bm{k} \cdot \bm{A} = 2 \pi .
        \label{eq:mcX001}
    \end{split}
\end{equation}
The Chern number of the other component $N_-$ is known to have the opposite sign and thus $N_-=-2\pi$. The contribution of this local model to $n_{{\cal M}_{(001)}}$ is $(N_+ - N_-)/(2\cdot 2\pi) = 1$. There are two X-points in the 1st BZ and it is easy to see that only the sign of the term with $M$ is different in the other X-point. Therefore, both nodal lines around the X-points have the same contribution to the mirror Chern number and thus the mirror Chern number is $n_{{\cal M}_{(001)}}=2$ when there are nodal lines on the $c k_z = 0$ plane (See table \ref{tab:phase}(C)(D)). It should be noted that in this calculation it did not matter whether some other nodal lines are touching the nodal line on the mirror plane. Especially in table \ref{tab:phase}(D), $n_{{\cal M}_{(001)}}$ is also 2.
A Berry curvature is also calculated from Eq.(\ref{eq:bconnX001}) as
\begin{equation}
    \begin{split}
        B_z = \frac{\chi^2}{2 R_+^3} & \left( (J+L) + K(k_x^2 + k_y^2) \phantom{\frac{Z}{Z}} \right. \\
        & \left. + M k_x k_y - 2M k_x k_y \frac{Z_+^2}{(R_+ + Z_+)^2} \right).
        \label{eq:bcurv001}
    \end{split}
\end{equation}
Because the mirror Chern number is a topological invariant, even in small $\chi$ case the mirror Chern number must be kept. The last term in Eq.(\ref{eq:bcurv001}) goes 0 when it is integrated. The $(k_x,k_y)$ dependence of the first three terms for a small $\chi$ is shown in Fig.\ref{fig:bcurv}(a). The $k_x$ and $k_y$ are normalized by $k_0 = \sqrt{|J+L|/K}$. The Berry curvature has a sharp ridge on the line where the nodal line exists when SOC is neglected. From this fact, the nodal line can be considered as a source of the mirror Chern number.

We move to the mirror Chern number $n_{{\cal M}_{(1\bar{1}0)}}$. The representation of the mirror operation $(x,y,z)\to(y,x,z)$ is now
\begin{equation}
    \sigma_0 \frac{1}{\sqrt{2}} (-i s_x + i s_y) = \left(
    \begin{array}{cccc}
        0 & 0 & e^{-i\frac{\pi}{4}} & 0 \\
        0 & 0 & 0 & e^{-i\frac{\pi}{4}} \\
        -e^{i\frac{\pi}{4}} & 0 & 0 & 0 \\
        0 & -e^{i\frac{\pi}{4}} & 0 & 0 \\
    \end{array} \right).
\end{equation}
To block diagonalize the mirror operator, a unitary transformation
\begin{equation}
    U = \frac{1}{\sqrt{2}} \left(
    \begin{array}{cccc}
        -e^{i\frac{\pi}{4}} & 0 & e^{i\frac{\pi}{4}} & 0 \\
        0 & -e^{i\frac{\pi}{4}} & 0 & e^{i\frac{\pi}{4}} \\
        1 & 0 & 1 & 0 \\
        0 & 1 & 0 & 1 
    \end{array} \right)
\end{equation} is defined and it acts as
\begin{equation}
    U^\dagger \sigma_0 \frac{1}{\sqrt{2}} (-i s_x + i s_y) U = \left(
    \begin{array}{cccc}
    i & 0 & 0 & 0 \\
    0 & i & 0 & 0 \\
    0 & 0 & -i & 0 \\
    0 & 0 & 0 & -i 
    \end{array} \right).
\end{equation}
The tight-binding model is transformed as
\begin{equation}
    \begin{split}
        & U^\dagger H_{4 \times 4}(\bm{k}) U \\
        &= \left(
        \begin{array}{cccc}
        Z & -iY + \chi k_d & 0 & i \chi k_{\perp} \\
        iY + \chi k_d & -Z & i \chi k_{\perp} & 0 \\
        0 & -i \chi k_{\perp} & Z & -iY - \chi k_d \\
        -i \chi k_{\perp} & 0 & iY - \chi k_d & -Z
        \end{array} \right) ,
    \end{split}
    \label{eq:socX110}
\end{equation}
\begin{equation}
    \begin{split}
        k_d &= \frac{1}{\sqrt{2}} (k_x + k_y) ,\\
        k_{\perp} &= \frac{1}{\sqrt{2}} (k_x - k_y).
    \end{split}
\end{equation}
On the mirror invariant plane, the $k_x=k_y$ plane, $k_{\perp}=0$ and the Eq.(\ref{eq:socX110}) is block digonalized. The $+i$ component is written as
\begin{equation}
    H_+(\bm{k})= X_+ \sigma_x + Y_+ \sigma_y + Z_+ \sigma_z ,
\end{equation}
\begin{equation}
    \begin{split}
        X_+ &= \chi k_d ,\\
        Y_+ &= Y \simeq V k_z ,\\
        Z_+ &= Z \simeq (J+L) + (K + \frac{M}{2})k_d^2 ,
    \end{split}
\end{equation}
where $V$ is defined as
\begin{equation}
    V = \left\{
    \begin{array}{l}
        c(S + \frac{T}{4}k_d^2) ~~~ (ck_z =0) ,\\
        c(S - \frac{T}{4}k_d^2) ~~~ (ck_z =\pi) .
    \end{array} \right.
\end{equation}
For $ck_z=0$, V is always positive. On the other hand, $V_{c k_z = \pi}$ is positive before the TR+inversion protected nodal line touches the (001) mirror plane ($\eta<0.2311$) but negative after it touches ($\eta>0.2311$).

The Berry connection for the occupied band of $H_+(\bm{k})$ is
\begin{equation}
    \begin{split}
        A_{+,d} &= \frac{\chi Y_+}{2 R_+ (R_+ - Z_+)} =\frac{ \chi V k_z}{2 R_+ (R_+ - Z_+)} ,\\
        A_{+,z} &= \frac{- V X_+}{2 R_+ (R_+ - Z_+)} = \frac{- \chi V k_d}{2R_+ (R_+ - Z_+)}. 
    \end{split}
\end{equation}
By integrating this along an ellipse written as $(k \cos \theta /\chi , k \sin \theta /|V|)$ in the $(k_d,k_z)$ plane and considering $k/\chi \to \infty, k/|V| \to \infty$, the Chern number of the $+i$ component $N_+$ is calculated for $V>0$ case as
\begin{equation}
    N_+ = \oint d \bm{k} \cdot \bm{A} = - 2 \pi .
    \label{eq:mcX110}
\end{equation}
As explained before, $N_- = 2\pi$ and the contribution to the mirror Chern number $n_{{\cal M}_{(1\bar{1}0)}}$ is -2 in this model. When $V<0$, it is easy to see that the sign of the Chern number get opposite and the contribution to $n_{{\cal M}_{(1\bar{1}0)}}$ is 2. The Berry curvature is written as
\begin{equation}
    B_{\perp} = - \frac{\chi V}{2 R_+^3} \left( |J+L| + (K + \frac{M}{2})k_d^2 \right).
\end{equation}
The $(k_x,k_y)$ dependence of $|B_{\perp}|$ for small $\chi$ is shown in Fig.\ref{fig:bcurv}(b). The $k_x$ and $k_y$ are normalized by $k_0=\sqrt{|J+L|/(K+M/2)}$. The $B_{\perp}$ has sharp peaks on the points where the nodal line penetrates the mirror plane when SOC is neglected. Similarly to the former case, the nodal line can be considered as a source of the mirror Chern number.

\section{\label{app:CNP} Calculation of the mirror Chern number (P-point)}

Next, let us focus on the $n_{{\cal M}_{(1\bar{1}0)}}$ around the P-point. Expanding Eq.(\ref{eq:socX110}) around $c k_z = \pi$, the $+i$ component is written as
\begin{equation}
    H_+(\bm{k}) = X_+ \sigma_x + Y_+ \sigma_y + Z_+ \sigma_z ,
\end{equation}
\begin{equation}
    \begin{split}
        X_+ &= \chi k_d ,\\
        Y_+ &= Y \simeq V k_z ,\\
        Z_+ &= Z \simeq (J-L) + K k_d^2 ,
    \end{split}
\end{equation}
where V is defined as
\begin{equation}
    V = c (S \cos c k_z + \frac{T}{4} k_d^2 \cos \frac{c k_z}{2}).
\end{equation}
The nodes are always in $\pi < c k_z < 2\pi$ and $V$ is always negative in that region.
The Berry curvature and $N_+$ are calculated as
\begin{equation}
    \begin{split}
        A_{+,d} &= \frac{- \chi Y_+}{2 R_+ (R_+ - Z_+)} =\frac{-\chi |V| k_z}{2 R_+ (R_+ - Z_+)} ,\\
        A_{+,z} &= \frac{-|V| X_+}{2 R_+ (R_+ - Z_+)} = \frac{ \chi |V| k_d}{2R_+ (R_+ - Z_+)}.
    \end{split}
\end{equation}
\begin{equation}
    N_+ = \oint d \bm{k} \cdot \bm{A} = 2 \pi .
    \label{eq:mcP110}
\end{equation}
There are also two P-points in 1st BZ and thus the contribution to the mirror Chern number $n_{{\cal M}_{(1\bar{1}0)}}$ is 2 in this model around the P-point.

\section{\label{sec:materials}Calculations for all combinations}
\subsection{Band dispersion}

\begin{figure*}
    \centering
    \includegraphics[width=15cm]{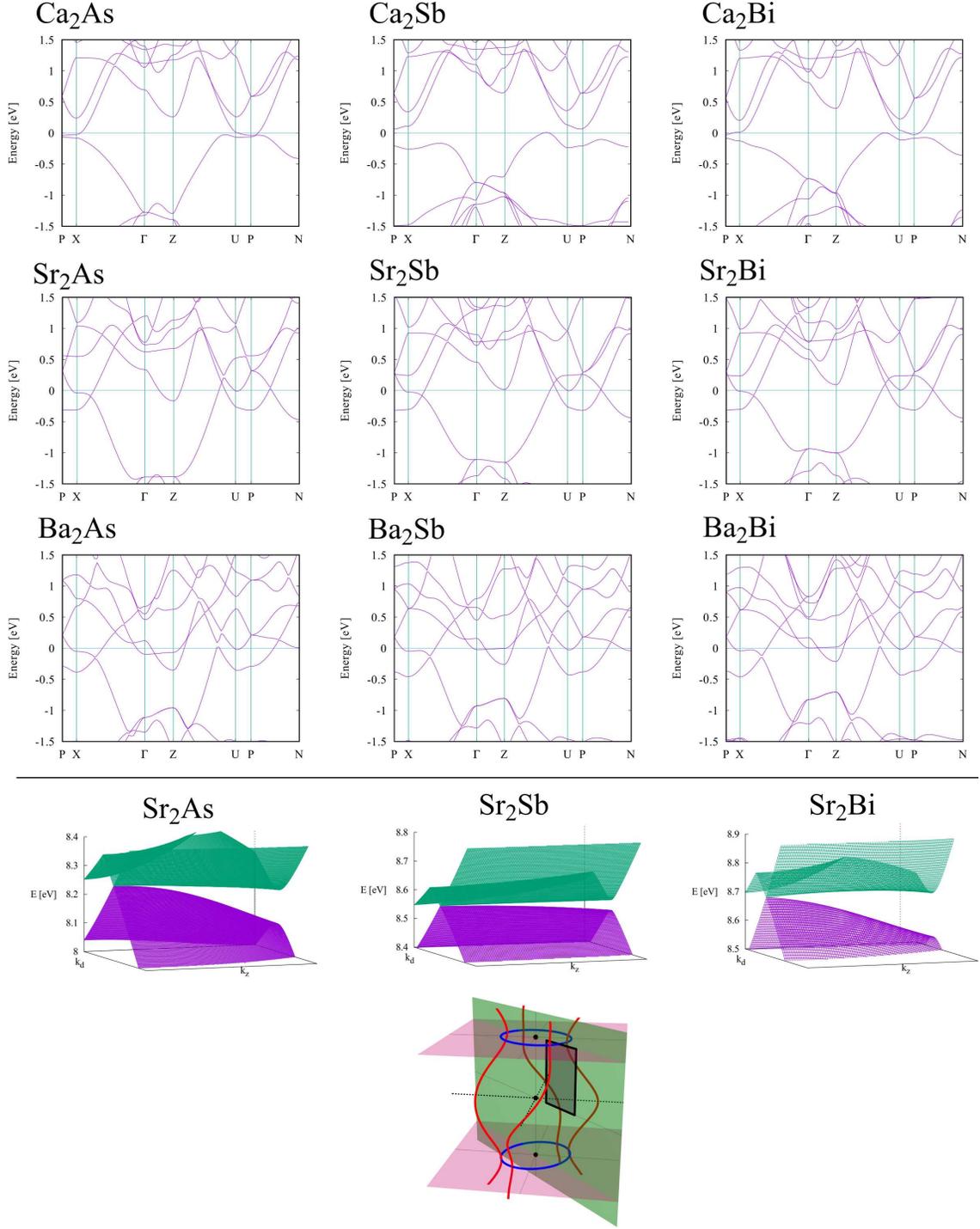}
    \caption{The band structures of all combination of atoms are shown. For Sr$_2Y$, the band structure around the mirror protected nodal line is also shown in the lower part of the picture. These band structure is calculated on the area shown by a shaded square in the schematic picture on the bottom.}
    \label{fig:allband}
\end{figure*}

 The band structures of all combination of atoms are shown in Fig.\ref{fig:allband}. For Sr$_2Y$, more detailed calculations is needed to know the configuration of nodal lines and the result is shown in the lower part of the Fig.\ref{fig:allband}. The detailed bands are calculated on the area shown by shaded square in the schematic picture at the bottom. In all of Sr$_2Y$, only one node is found in this area. This result means that the red nodal line dose not penetrate the green mirror plane. All of Ba$_2Y$ have larger overlap between the valence band and the conduction band compared to that of Sr$_2Y$. Considering the discussion on the tight-binding model, Ba$_2Y$ should belong to the phase (D) in table \ref{tab:phase}.
 
\subsection{Charge Density}
\begin{figure}
    \centering
    \includegraphics[width=8cm]{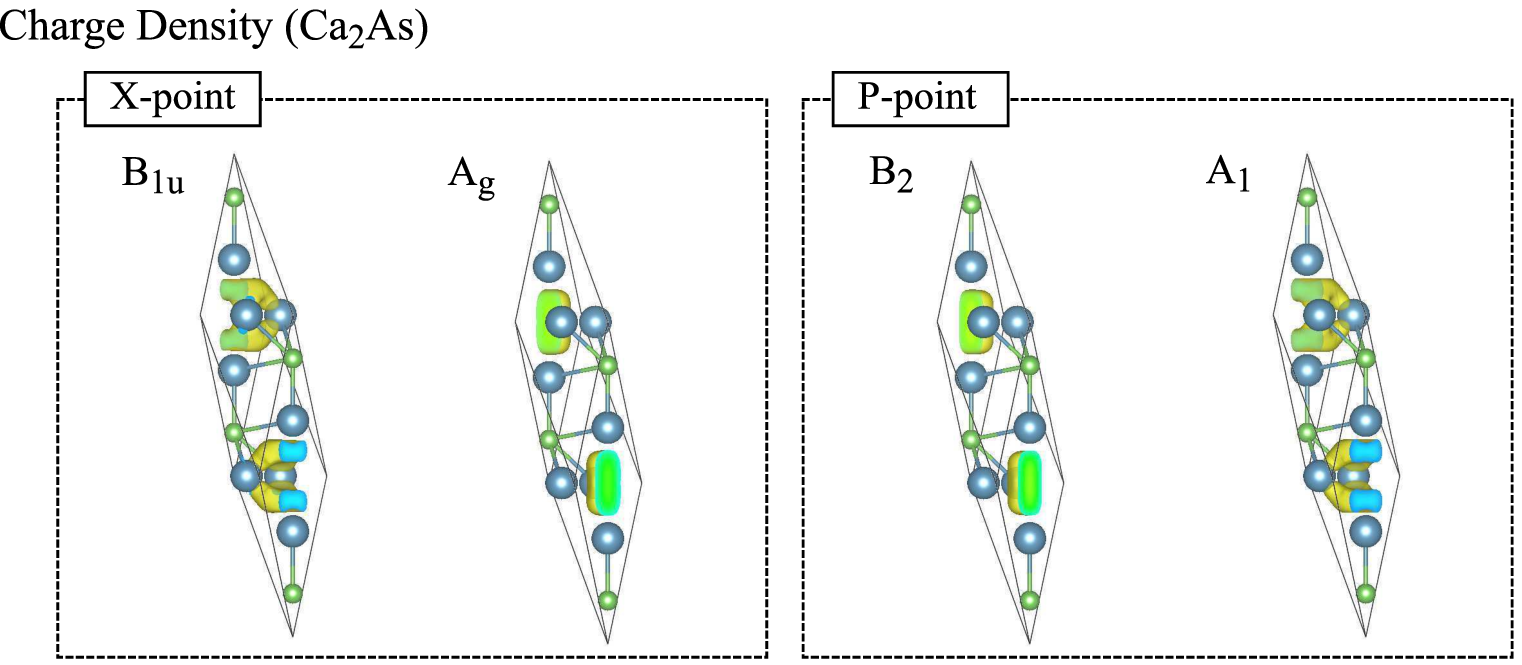}
    \caption{Charge density of the states in X- and P-point of Ca$_2$As for bands around the Fermi level. The electron charge is localized around the 2b position. $\mathrm{A_g}$ and $\mathrm{B_2}$ have no node structure, whereas the other two have a node on the bisector of two Ca(4e) atoms.}
    \label{fig:ChargeD}
\end{figure}

 As mentioned in Introduction, Sr$_2$Bi was suggested to be an electride with nodal lines in its bulk state \cite{Hirayama}, i. e., electrons are localized in the interstitial space between six Sr atoms, which is the Wyckoff position 2b (0,0,1/2).
 
 In Fig.\ref{fig:ChargeD}, charge densities of Ca$_2$As for each band in the P- and X-points are shown. This result shows that this system is a typical example of an electride, as in Sr$_2$Bi. Electrons in the states around the Fermi level are localized in an interstitial region between six Ca atoms. A$_g$ in the X-point and B$_2$ in the P-point have charge density distributions without node around the 2b position, which is like a bonding state. On the other hand, B$_{1u}$ in the X-point and A$_1$ in the P-point have charge density distributions with a node on a perpendicular bisector of two Ca(4e) atoms, which is like an anti-bonding state. We find that not only in Ca$_2$As but also in Ca$_2$Bi, Ca$_2$Sb, Sr$_2$As, Sr$_2$Sb, and Ba$_2Y$, the same charge densities are given to the states with the same irreps. Our results reveal that all materials in the Ca$_2$As family including Sr$_2$Bi are electrides.

\section{\label{sec:sstate}Surface State}
\begin{figure*}
    \centering
    \includegraphics[width=16cm]{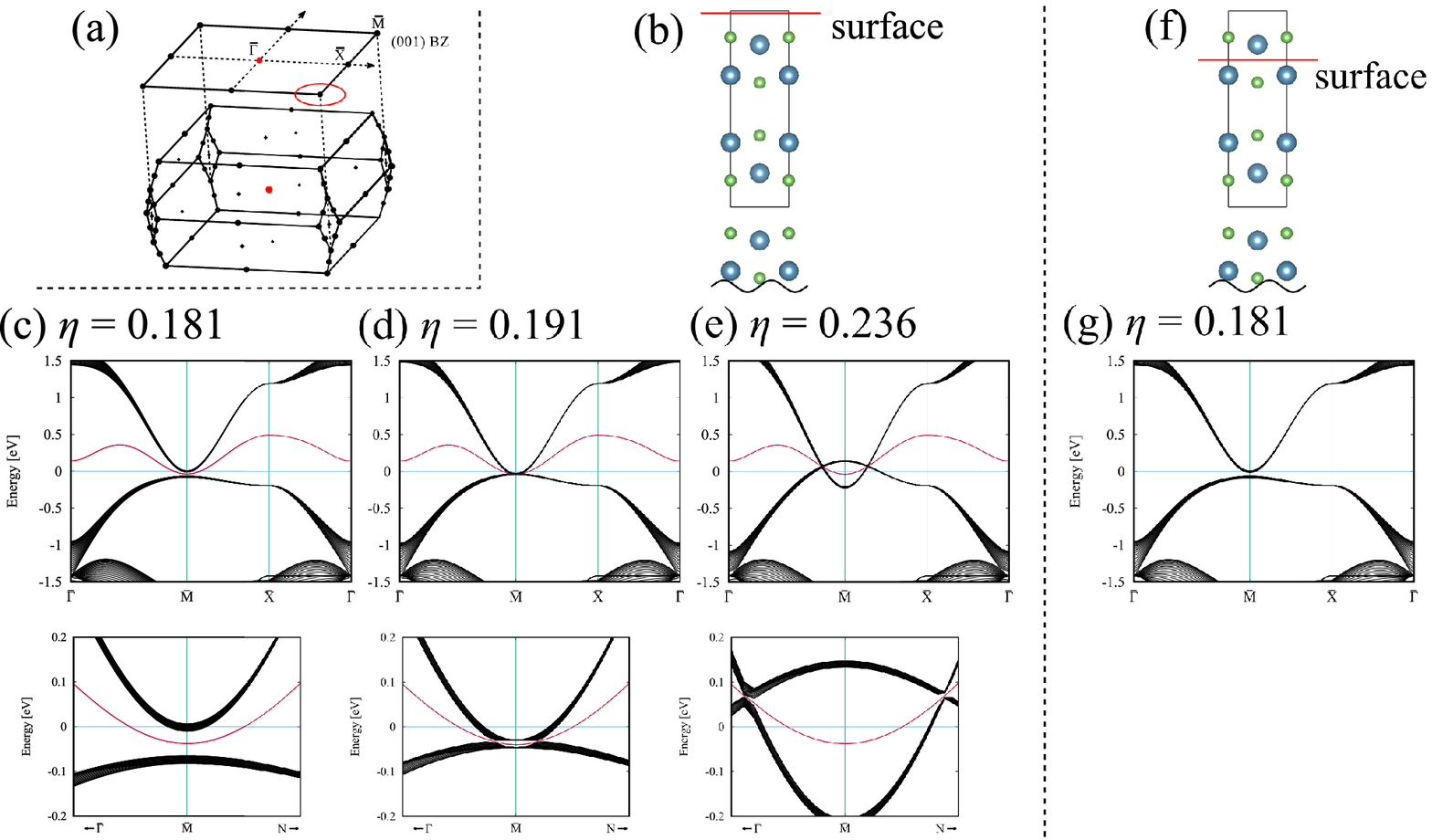}
    \caption{(a) The (001) surface BZ and the projected nodal line (a red ring). (b)(f) The red bar represents where the surface of the slab is located. The rectangle is the conventional cell. (c)(d)(e)(g) Surface band dispersions for each $\eta$. The red bands are surface states. In lower panel, magnified picture around the $\mathrm{\bar{M}}$ are shown. In (g) there is no surface state.}
    \label{fig:surstate}
\end{figure*}

 Here, we discuss the presence of surface states by slab calculation with the tight-binding model Eq.(\ref{eq:88tb}).
  Generally, a nodal line semimetal has a "drumhead-like" surface state inside (or outside) a projected nodal line on a surface BZ. In Ca$_2$As family, a projected nodal line appears roughly as a ring around $\bar{\mathrm{M}}$-point (Fig.\ref{fig:surstate}(a)) in both phases shown Fig.\ref{fig:tbband}(b) and Fig.\ref{fig:tbband}(c). All band dispersions are calculated with a slab of 15 conventional cells. Generally, surface states can be dependent on the configuration of the surface, thus we calculate two different surface configurations shown in Fig.\ref{fig:surstate}(b) and Fig.\ref{fig:surstate}(f). The band dispersions for three phases, Fig.\ref{fig:surstate}(c)(d)(e), are calculated with the surface configuration (b). All of them have surface state shown with red lines, even though there is no nodal line in the phase (c). In (c), the surface state is isolated with the bulk states and this is a typical surface state in electrides. The presence of the surface states is diagnosed by calculating the Zak phase. In the system with the inversion symmetry, the Zak phase $\omega_{\mathrm{Zak}}$ is easily calculated as a product of the inversion eigenvalue on some time-reversal invariant momenta (TRIM). 
 \begin{equation}
    \begin{split}
            \exp (i\omega_{\mathrm{Zak}}) &= \exp \left[ i \int_{\Gamma_1 \to \Gamma_0 \to \Gamma_1} d k_z A_z(k_z) \right] \\
            &= \prod_{\bm{k}=\Gamma_0,\Gamma_1} \prod_{n:\mathrm{occupied}} \xi_n(\bm{k})         
    \end{split}
 \end{equation}
\begin{equation*}
    \begin{split}
        \Gamma_0, \Gamma_1 &: \mathrm{TRIM~with~same~}(k_x,k_y) \\
        A_z (k_z) &: \mathrm{Berry~connection~in~}(k_x,k_y,k_z) \\
        \xi_n(\bm{k}) &: \mathrm{inversion~eigenvalue~of~n_{th}~band~in~} \bm{k}
    \end{split}
\end{equation*}
 To calculate the inversion eigenvalue, the inversion center must be taken to make the edge of the unit cell to be identical to the edge of the slab. In the case of Fig.\ref{fig:surstate}(b), the inversion center should be placed (1/4,1/4,1/4) in the fractional coordinate in the conventional cell. The products of the inversion eigenvalue for occupied bands are -1 on the $\mathrm{\Gamma}$-point and 1 on the Z-point. Therefore, the Zak phase on $\mathrm{\bar{\Gamma}}$-point on the surface BZ is $\pi$. $\pi$ Zak phase guarantees the presence of the surface state as long as the surface and the unit cell are correctly matching. Since the Zak phase is quantized, the Zak phase around the $\mathrm{\Gamma}$-point must be the same as that of the $\mathrm{\Gamma}$-point. The Zak phase can be changed only where the projected nodal line exists. However, in Ca$_2$As family system, two nodal lines around the P-point ($\pi$,$\pi$,$\pi/c$) and another P-point ($\pi$,$\pi$,$-\pi/c$) (the X-point ($\pi$,$\pi$,0) and another X-point ($\pi$,$\pi$,$2\pi/c$) for the phase Fig.\ref{fig:tbband}(c)). Therefore, the projected nodal line appears around the $\bar{\mathrm{M}}$-point is a doubly overlapped nodal line. Generally, a single projected nodal line changes the Zak phase by $\pi$. The doubly overlapped nodal line changes the Zak phase by $2\pi$ and thus the Zak phase, which is defined in mod 2, is the same inside and outside the projected nodal line. As a result, for the case of Fig.\ref{fig:surstate}(c)(d)(e), the Zak phase is $\pi$ and there are surface states everywhere in the surface BZ. The presence of the surface state is explained with a simple physical picture. It has been shown that an electron is localized in the 2b position and the position 2b is lying on the slab surface shown in  Fig.\ref{fig:surstate}(b). It means that the localized state is divided when the surface is made and the remained "half" appears as the surface state. 
 
 A band dispersion of a slab with the surface configuration Fig.\ref{fig:surstate}(f), in which "a layer" was removed from Fig.\ref{fig:surstate}(b), is shown in Fig\ref{fig:surstate}(g). For this surface configuration, the inversion center should be placed (1/2,1/2,1/2) in the fractional coordinate in the conventional cell. Because the inversion center is changed, the products of the inversion eigenvalue are changed form former case and now they are +1 both in the $\mathrm{\Gamma}$- and Z-point. In this case, the Zak phase is 0 and no surface state appears. Contrary to the surface configuration (b), the configuration (f) does not cut the state localized in the 2b position. The absence of the surface state, in this case, is consistent with the above explanation.


\end{document}